\documentclass[aps, prb, reprint, amsmath, amssymb,nobibnotes, tbtags]{revtex4-2}
\usepackage[colorlinks, allcolors=blue]{hyperref}

\usepackage{siunitx}
\usepackage{bm}
\usepackage{graphicx}
\usepackage[utf8]{inputenc}
\usepackage{mathtools}

\renewcommand{\vec}{\bm} 
\renewcommand{\Im}{\text{Im}}
\renewcommand{\Re}{\text{Re}}

\newcommand{\iu}{\mathrm{i}}
\newcommand{\e}{\text{e}}

\newcommand{\ud}{{ \uparrow \downarrow}}
\newcommand{\du}{{  \downarrow \uparrow}}
\newcommand{\uu}{{ \uparrow \uparrow}}
\newcommand{\dd}{{ \downarrow \downarrow}}

\newcommand{\eq}{Eq.\ } 
\newcommand{\tr}[1]{\text{Tr} \left[ #1 \right]}

\begin{document}

\title{Quantum Theory of Spin-Transfer and Spin-Pumping in Collinear Antiferromagnets and Ferrimagnets}
\author{Hans Gløckner Giil and Arne Brataas}
\affiliation{Center for Quantum Spintronics, Department of Physics, Norwegian University of Science and Technology, NO-7491 Trondheim, Norway}
\date{\today}

\begin{abstract}
Antiferromagnets are promising candidates as active components in spintronic applications. They share features with ferrimagnets in that opposing spin orientations exist in two or more sublattices. Spin transfer torque and spin pumping are essential ingredients in antiferromagnetic and ferrimagnet spintronics. This paper develops an out-of-equilibrium quantum theory of the spin dynamics of collinear magnets containing many spins coupled to normal metal reservoirs. At equilibrium, the spins are parallel or antiparallel to the easy axis. The theory, therefore, covers collinear antiferromagnets and ferrimagnets. We focus on the resulting semi-classical spin dynamics. The dissipation in the spin dynamics is enhanced due to spin-pumping. Spin accumulations in the normal metals induce deterministic spin-transfer torques on the magnet. Additionally, each electron's discrete spin angular momentum causes stochastic fluctuating torques on the antiferromagnet or ferrimagnet. We derive these fluctuating torques. The fluctuation-dissipation theorem holds at high temperatures, including the effects of spin-pumping.
At low temperatures, we derive shot noise contributions to the fluctuations.
\end{abstract}

\date{\today}

\maketitle
\section{Introduction}
\label{section:introduction}

Spin transfer torque (STT) and spin pumping (SP) are essential ingredients in the generation and detection of spin currents and are central components in modern spintronics research and devices~\cite{hirohataReviewSpintronicsPrinciples2020b}.
The use of magnetic insulators enables signal propagation without moving charges and could provide low-dissipation and ultra-fast memory devices~\cite{Brataas:PhysRep2020}.
Initially, much of spintronic research focused on the study of STT~\cite{Berger:PRB1996, slonczewskiCurrentdrivenExcitationMagnetic1996a, brataasFiniteElementTheoryTransport2000} and SP~\cite{Tserkovnyak:PRL2002, heinrichDynamicExchangeCoupling2003, tserkovnyakNonlocalMagnetizationDynamics2005c} in ferromagnets (FMs). Subsequently, this included also works on fluctuations~\cite{foros_magnetization_2005, Chudnovskiy:PRL2008, Swiebodzinski:PRB2010} and pumped magnon condensates~\cite{Bender2012, Bender2014, divinskiyEvidenceSpinCurrent2021}.

Unlike FMs, whose macroscopically apparent magnetic properties have been known for thousands of years, antiferromagnets (AFMs) carry zero net magnetic moments and were elusive for some time.
Even after their discovery, AFMs were believed to have few potential applications~\cite{neel_nobel_1970} and were disregarded in the early days of spintronics research.
Recent theoretical and experimental findings have highlighted the potential of using AFMs in spintronics applications, thus starting the field of antiferromagnetic spintronics. Key discoveries were the robustness of AFMs to external magnetic perturbations and the high resonance frequency of antiferromagnetic material~\cite{jungwirth_antiferromagnetic_2016, Rezende:JAP2019}.
The prediction \cite{nunez_prb_2006} and subsequent experimental detection \cite{wei_changing_2007} of an STT in AFMs sparked a massive interest in using AFMs as the active component in spintronics devices~\cite{jungwirth_antiferromagnetic_2016, baltz_antiferromagnetic_2018}.
Moreover, it was predicted that contrary to what was believed, antiferromagnets are as efficient in pumping spin currents as FMs~\cite{chengSpinPumpingSpinTransfer2014a}.
This effect was later experimentally detected in the easy-axis AFM MnF$_2$~\cite{Vaidya:Science2020}.
These discoveries opened up the possibility of utilizing AFMs in spintronic applications, enabling the possible fabrication of stray-field-free devices operating in the \si{\tera \hertz}-regime~\cite{jungwirth_antiferromagnetic_2016, keffer_theory_1952}, allowing for much faster device operation than in FMs.

In recent years, the spin dynamics in AFMs have been explored extensively, including the effects of disorder \cite{Gulbrandsen:PRB2018}, generation of spin-Hall voltages~\cite{johansenSpinPumpingInverse2017a}, and the properties of antiferromagnetic skyrmions \cite{barkerStaticDynamicalProperties2016}.
The spin dynamics of ferrimagnetic materials have also been studied~\cite{kamraSpinPumpingShot2017a}. Phenomenological models of intra- and cross-lattice torques were introduced in \cite{kamraGilbertDampingPhenomenology2018a}. Ref.\ \onlinecite{tangAbsenceCrosssublatticeSpin2023c} further discusses the competition between intra and cross-sublattice spin pumping in specific models of antiferromagnets.

As in antiferromagnets, ferrimagnets have opposing magnetic moments. However, these moments have different magnitudes, resulting in a net magnetization. These features result in rich spin dynamics ranging from behavior reminiscent of antiferromagnets to ferromagnets. A prime example of a ferrimagnet is yttrium-iron-garnet (YIG). The low-energy magnon modes in YIG resemble modes in ferromagnets.

In the study of non-equilibrium effects, the Keldysh path integral approach to non-equilibrium quantum field theory is a powerful tool in the study of non-equilibrium systems beyond linear response~\cite{kamenev_field_2011,keldysh_jetp_1964}. 
Although most of the research on STT and SP utilized a semiclassical approach, some works have used the Keldysh framework in the study of spin dynamics in FMs out of equilibrium~\cite{Chudnovskiy:PRL2008, Swiebodzinski:PRB2010, Takei:PRB2019, duineFunctionalKeldyshTheory2007, tassiQuantumNoiseSpin2019}. Moreover, the Keldysh method was recently used to formalize a fully quantum mechanical theory of STT and SP, including the effects of quantum fluctuations~\cite{Brataas:PRB2022}. 
These fluctuations have become increasingly relevant with the development of new devices operating in the low-temperature regime. 
Nevertheless, applying the Keldysh method to derive microscopic relations for SP, STT, and fluctuating torques in an AFM system is lacking.

In this paper, we extend the approach of Ref.~\onlinecite{Brataas:PRB2022}, which examined a ferromagnet in the macrospin approximation coupled to normal metals featuring spin and charge accumulation, to a similar system but instead featuring a collinear magnet with many individual spins coupled at different sublattices to normal metals. Our study thus covers antiferromagnets, ferrimagnets, and ferromagnets. We derive the spin dynamics using a fully quantum mechanical Keldysh non-equilibrium approach. We find expressions for the spin transfer torque, spin-pumping-induced Gilbert damping, and fluctuating fields, including low-temperature shot-noise contributions. The Gilbert damping and fluctuations contain both inter-lattice and intra-lattice terms. Using Onsager reciprocal relations, we relate the spin pumping and spin transfer coefficients. 
Our results enhance the knowledge of the microscopic expressions of STT and SP and fluctuating torques in antiferromagnets and ferrimagnets coupled to normal metals in the low-energy regime, where quantum fluctuations become essential.

The subsequent sections of this paper are structured as follows.
In Sec.~\ref{section:model}, we introduce the model employed for the itinerant electrons in the normal metals, the localized magnetic moments in the antiferromagnetic or ferrimagnet, and the electron-magnon coupling between them. 
We then present the key findings of this paper in Sec.~\ref{section:main_results}, including microscopic definitions of the spin transfer torque, spin pumping, and fluctuating torques in many spin magnets, being an antiferromagnetic, ferrimagnet, or ferromagnet. The derivation of an effective magnon action, achieved by integrating fermionic degrees of freedom resulting from the interaction with normal metals, is detailed in Sec.~\ref{section:effective_action}. The evaluation of this effective action is then provided in Sec.~\ref{section:spindynamics}.
Finally, Sec.~\ref{section:conclusion} concludes the paper.

\section{Model}
\label{section:model}
We consider a bipartite collinear magnet coupled to an arbitrary number of normal metal reservoirs. The magnet can represent an antiferromagnet, a ferrimagnet, or a ferromagnet. 
The total Hamiltonian is 
\begin{equation}
    \hat{H}  =  \hat{H} _e + \hat{H} _{em} + \hat{H} _m 
\end{equation}
in terms of the Hamiltonian describing the electrons in the normal metal $\hat{H} _e$, the Hamiltonian describing the interaction between the electrons and the magnet $\hat{H} _{em}$, and the Hamiltonian of the magnet $\hat{H} _m$. 

The Hamiltonian of the electrons combined with the Hamiltonian representing the interaction between the electrons and the magnet is 
\begin{equation}
    \hat{H}_e + \hat{H}_{em} =  \int d{\bf r} \hat{\psi}^\dag  \left[ H_e+ \hbar^{-1} \sum_i u_i\bm{\sigma} \cdot \hat{\bm S}_i \right] \hat{\psi}\, , 
\label{H}
\end{equation}
where $\hat{\psi}^\dag=( \hat{\psi}_\uparrow^\dag, \hat{\psi}_\downarrow^\dag )$ is the spatially dependent 2-component itinerant electron field operator, and ${\bm \sigma}$ is the vector of Pauli matrices in the $2 \times 2$ spin space. 
In the Hamiltonian \eqref{H}, $u_i (\vec r)$ represents the spatially dependent exchange interaction between the localized spin at site $i$ and the itinerant electrons. This interaction is localized around spin $i$ inside the magnet.
The sum over the localized spins $i$ consists of a sum over sites in sublattice $\mathcal A$ and sublattice $\mathcal B$, i.e., $\sum_i \ldots \rightarrow \sum_a \ldots +  \sum_b \ldots $. 
The localized spin operator $\hat{\bm S}_i$ has a total spin angular momentum $S_i = \hbar \sqrt{s_i(s_i + 1)}$ where $s_i$ is the (unitless) spin quantum number of the localized spin, such that $\hat{\bm S_i}^2= \hbar^2 s_i(s_i+1)$. For large $s_i$, the difference between $S_i / \hbar$ and $s_i$ is a first-order correction, and we can approximate $S_i \approx \hbar s_i$.

The spin-independent part of the single-particle electron Hamiltonian is
\begin{equation}
    H_e = - \frac{\hbar^2}{2m} \nabla^2 + V_c \, , 
    \label{H_0}
\end{equation}
where $V_c$ is the spatially dependent charge potential. 

In the classical limit of the magnet, the spins at sublattice $\mathcal A$ are along a certain direction and the spins at sublattice $\mathcal B$ are along the opposite direction in the ground state. We will consider the semiclassical spin dynamics near the instantaneous classical direction of the spins that we let be along the $z$ direction and adiabatically adjust the evolution of the small deviation \cite{Chudnovskiy:PRL2008, Swiebodzinski:PRB2010, Brataas:PRB2022}.
In the following, it is constructive to expand the interaction term to the second order in the magnet creation/annihilation operators using a Holstein-Primakoff transformation,
\begin{equation}
    \hat H_{em} = \hat H_0 + \hat H_1 + \hat H_2 \, ,
\end{equation}
where $\hat H_0$ is the interaction with the classical magnetic ground state and $\hat H_1$ ($\hat H_2$) is the interaction term to the first (second) order.
The classical ground state contribution to the interaction is then
\begin{equation}
    \hat{H}_{0} =  \int d{\bf r} \hat{\psi}^\dag  V_s \sigma_z  \hat{\psi}\, , 
\end{equation}
where the magnitude of the spatially dependent spin potential experienced by the itinerant electrons is  
\begin{align}
    V_s ({\bm r})=\sum_a s_a u_{a} ({\bm r}) - \sum_b  s_b u_{b} ({\bm r}) \, ,
\end{align}
and oscillates rapidly with the staggered field.

In the macrospin approximation, $\sum_i \vec S_i$ can be treated as a giant spin in ferromagnets. Then, $u_{i}(\vec r)$ becomes the effective exchange interaction.
Ref.\ \onlinecite{Brataas:PRB2022} shows how the electronic Hamiltonian $\hat H_e$ combined with the electron-magnon Hamiltonian to zeroth order $\hat H_0$ become particularly transparent in ferromagnet-normal metal systems in terms of the scattering states of the itinerant electrons for the macrospin dynamics.
We generalize this approach to magnet-normal metal systems with individual localized spins. 
In this picture, the electronic Hamiltonian remains simple, as in Ref.\ \onlinecite{Brataas:PRB2022}:
\begin{equation}
    \hat{H}_e + \hat H_0 = \sum_{s\alpha} \epsilon_\alpha \hat{c}_{s\alpha}^\dag \hat{c}_{s\alpha}  \, , 
\label{H_escat}
\end{equation}  
where $\hat{c}_{s\alpha}$ annihilates an electron with spin $s$ ($s=\uparrow$ or $s=\downarrow$). The quantum number $\alpha=\kappa n\epsilon$ captures the lead $\kappa$, the transverse waveguide mode $n$, and the electron energy $\epsilon$. The electron energy consists of a transverse contribution $\epsilon_n$ and a longitudinal contribution $\epsilon(k)=k^2/2m$, where $k$ is the longitudinal momentum, such that $\epsilon=\epsilon_n + \epsilon(k)$. The eigenenergy is spin degenerate, since the leads are paramagnetic. Furthermore, we consider identical leads such that the eigenenergy is independent of the lead index.
\begin{figure}[htb]
    \centering
    \includegraphics[width = 0.47 \textwidth]{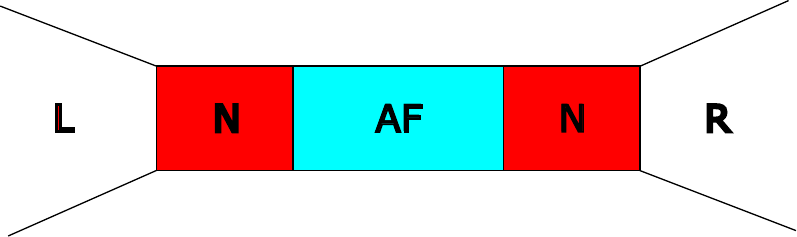}
    \caption{An antiferromagnet (AF) with conductors (N) on either side connected to
    a right (R) and a left (L) lead.}
    \label{fig:system_setup}
\end{figure}  
The system setup is shown in Fig.~\ref{fig:system_setup} for the case of two leads.
In \eq \eqref{H_escat} and similar expressions to follow, the sum over the scattering states implies that $\sum_\alpha X_{s\alpha}$ = $\sum_{\kappa n} \int_{\epsilon_n}^\infty d\epsilon  X_{s\kappa n}(\epsilon)$.
In the scattering approach, the field operator is 
\begin{equation}
    \hat \psi_s = \sum_{ \alpha} \hat c_{s \alpha} \psi_{s \alpha}\, ,
\end{equation}
where $\psi_{s \alpha}(\vec r)$ is the wave function of a scattering state of spin $s$ and quantum number $\alpha$.

The Hamiltonian of the antiferromagnet is [$i$ denotes a site at sublattice $\mathcal A$ ($i=a$) or $\mathcal B$ ($i=b$)]
\begin{align}
    H_m  =&  \hbar^{-2} \sum_{ij} J_{ij} \hat{\bf S}_i \cdot \hat {\bf S}_j - K  \hbar^{-2} \sum_i  \left( \hat{\bf S}_i \cdot \vec z \right)^2 \nonumber \\ 
    & + \gamma \mu_0  \sum_a \vec H^A_a \cdot \hat{\vec S_{a}} 
    + \gamma \mu_0  \sum_b \vec H^B_b \cdot \hat{\vec S_{b}} \, , 
    \label{Hm}
\end{align}
where $J_{ij}$ is the symmetric exchange interaction, $K>0$ is the easy-axis anisotropy energy, and $\gamma = g^* \mu_B / \hbar$ is the (absolute value of) the effective gyromagnetic ratio, where $g^*$ is the effective Landé g-factor and $\mu_B$ is the Bohr magneton.
In \eq \eqref{Hm}, $\vec H^{A, B}_{i}$ is the external magnetic field in units of \si{\ampere  \meter^{-1}} at lattice site $i = \{a, b\}$, and $\mu_0$ is the vacuum permeability, which appears because we are employing SI units.
In reality, $\vec H^A = \vec H^B$ in the presence of a uniform external magnetic field. 
However, to illustrate and understand the physics, we allow the external fields at sublattices $\mathcal A$ and $\mathcal B$ to differ, and to depend on the lattice site.

We consider the low-energy excitations from the semiclassical ground state of the staggered spin orientation. To this end, we carry out a Holstein-Primakoff expansion to the second order in magnon excitations at each sublattice $\mathcal A$ and $\mathcal B$ described via the annihilation operators $\hat{a}_a$ and $\hat{b}_b$ as detailed in Appendix~\ref{appendix:transformations}. 
Introducing the raising/lowering fields as $H_{\pm} = H_{x} \pm \iu H_{y}$, the magnon Hamiltonian becomes
\begin{align}
    H_m  =&   E_0 + \sum_{a} E^A_{a} \hat  a^\dagger_a \hat a_a  + \sum_{b} E^B_{b} \hat b^\dagger_b \hat b_b
    \nonumber
    \\
    &
    +
    2 \sum_{aa'}J_{aa'} \sqrt{s_a s_{a'}} \hat{a}^\dagger_a \hat a_{a'} 
    +2 \sum_{bb'}J_{bb'} \sqrt{s_b s_{b'}} \hat b^\dagger_b \hat b_{b'}
    \nonumber
    \\
    &
    +
    2 \sum_{ab}J_{ab} \sqrt{s_a s_{b}} [\hat{a}_a \hat b_b +\hat a^\dagger_a \hat b^\dagger_b] 
    \nonumber
    \\
    &
    +  \gamma \mu_0 \hbar \sum_a \sqrt{\frac{s_a}{2}} [ H^A_{a-} \hat a_a + H^A_{a+} \hat a_a^\dagger]
    \nonumber
    \\
    &
    +  \gamma \mu_0 \hbar \sum_b \sqrt{\frac{s_b}{2}} [ H^B_{b-} \hat b^\dagger_b + H^B_{b+} \hat b_b ]\, ,
    \label{Hm_HP}
\end{align}
where the classical ground state energy $E_0$ is
\begin{align}
    \nonumber
    E_0 & =  \sum_{aa'} s_a s_{a'} J_{aa'} +  \sum_{bb'} s_b s_{b'} J_{bb'} - 2 \sum_{ab} s_a s_{b}J_{ab}
    \\
    &- 2 K \sum_i s_i^2 + \hbar   \mu_0 \sum_{a} s_a H^A_{az} -  \hbar   \mu_0 \sum_{b} s_b H^B_{bz} \, ,
\end{align}
and is disregarded in the following. 
\begin{align}
\nonumber
    E^{A(B)}_{a(b)} &= 2  \sum_{b (a)} s_{b (a)} J_{ab} 
    - \sum_{a'(b')}
    s_{a'(b')}J_{a(b)a'(b')} 
    \\
    &+ 2 s_{a(b)} K \mp \hbar \gamma \mu_0 H^{A (B)}_{a(b)z}
\end{align}
is the energy of a local excitation, where the upper sign holds for sites on sublattice $\mathcal A $ and the lower sign holds for sites on sublattice $\mathcal B$.

In the scattering basis of the electronic states, the corrections to the antiferromagnetic ground state electron-magnon interaction to quadratic order in the magnet operators becomes
$\hat{H}_{em} - \hat H_0 = \hat{H}_{1} + \hat{H}_2$.
The first-order contribution of electron-magnon interaction is
\begin{align}
    \hat{H}_{1} =  & \sum_{a \alpha \beta} \sqrt{\frac{2}{s_a}}  \left[  \hat{a}_a  \hat{c}_{\downarrow \alpha}^\dag  W_{a \downarrow \uparrow}^{\alpha \beta}\hat{c}_{\uparrow \beta} + \hat{a}_a^\dag  \hat{c}_{\uparrow \alpha}^\dag  W_{a\uparrow \downarrow}^{\alpha \beta}\hat{c}_{\downarrow \beta} \right] \nonumber \\
    & + \sum_{b\alpha \beta} \sqrt{\frac{2}{s_b}}  \left[  \hat{b}_b^\dag  \hat{c}_{\downarrow \alpha}^\dag  W_{b \downarrow \uparrow}^{\alpha \beta}\hat{c}_{\uparrow \beta} + \hat{b}_b  \hat{c}_{\uparrow \alpha}^\dag  W_{b\uparrow \downarrow}^{\alpha \beta}\hat{c}_{\downarrow \beta} \right] \, ,
    \label{H1}
\end{align}
and describes the spin-flip scattering of the itinerant electrons associated with creating or annihilating localized magnons. 
The dimensionless matrix $W_i$ is governed by the exchange potential $u_{i}({\bm r})$ and the scattering states wave functions $\psi_{s\alpha}$:
\begin{equation}
    W_{i s s`}^{\alpha \beta} =  \int d\vec r \psi^*_{s \alpha}(\vec r)   s_i u_{i}(\vec r) \psi_{s` \beta}(\vec r) \, , 
    \label{W}
\end{equation}
and is Hermitian, $W_{i\uparrow \downarrow}^{\alpha \beta} = [W_{i\downarrow \uparrow }^{ \beta \alpha}]^* $. 
The electron-magnon interaction that is second order in the magnon operators is 
\begin{align}
\nonumber
    \hat{H}_{2}  =& -  \sum_{a\alpha \beta} \frac{\hat{a}_a^\dag \hat{a}_a }{s_a}   \left[ \hat{c}_{\uparrow \alpha}^\dag  W_{a\uparrow \uparrow}^{\alpha \beta} \hat{c}_{\uparrow \beta} - \hat{c}_{\downarrow \alpha}^\dag W^{ \alpha \beta}_{a \downarrow \downarrow} \hat{c}_{\downarrow \beta}  \right] \\
    & + \sum_{b\alpha \beta}  \frac{\hat{b}_b^\dag \hat{b}_b}{s_b}   \left[ \hat{c}_{\uparrow \alpha}^\dag  W_{b \uparrow \uparrow}^{\alpha \beta} \hat{c}_{\uparrow \beta} - \hat{c}_{\downarrow \alpha}^\dag W^{ \alpha \beta}_{b \downarrow \downarrow} \hat{c}_{\downarrow \beta}  \right] \, , 
    \label{H2}
\end{align}
where the matrix elements are defined in \eq \eqref{W}.
We note that our electron-magnon-interaction is isotropic in spin space and will give rise to zeroth-, first-, and second-order magnon terms in the Hamiltonian, i.e. $\hat H_0$, $\hat H_1$ and $\hat H_2$, respectively. This is in contrast to the model used in Refs.\ \onlinecite{Takei:PRB2019, Bender2012}, where only the first-order term $\hat H_1$ is considered.

Finally, in normal metal reservoirs, the occupation of the state is 
\begin{equation}
    \label{eq:occupation}
    \langle c_{s'\alpha}^\dag c_{s \beta} \rangle = \delta_{\alpha \beta} n_{s s' \alpha} \, ,
\end{equation}
where the $2\times 2$ out-of-equilibrium distribution is
\begin{align}
\nonumber
n_{s s' \alpha}   =& \frac{1}{2} \left[ f_{\kappa \uparrow}(\epsilon_\alpha)) + f_{\kappa \downarrow}(\epsilon_\alpha)) \right] \delta_{s s'} \\
& + \frac{1}{2} \left[ f_{\kappa \uparrow}(\epsilon_\alpha)) - f_{\kappa \downarrow}(\epsilon_\alpha)) \right] {\bf u}_\kappa \cdot \vec \sigma_{ss'} \, ,
\label{n_ssa}
\end{align} 
allowing for a (lead-dependent) spin accumulation in the direction of the unit vector $\vec u_k$. 
$f_\uparrow$ and $f_\downarrow$ are general distribution functions for spin-up and spin-down particles, which generally differ for elastic or inelastic transport~\cite{Brataas:PRB2022}. In equilibrium, the distribution function only depends on energy,  
\begin{align}
    f_{\kappa \uparrow}^{\text{eq}}(\epsilon) = f_{\kappa \downarrow}^{\text{eq}}(\epsilon) = f(\epsilon - \mu_0) \, ,
\end{align}
where $f$ is the equilibrium Fermi-Dirac distribution and $\mu_0$ is the equilibrium chemical potential.

In inelastic transport, the spin- and charge accumulations $\mu^C$ and $\mu^S$ correspond to chemical potential in a (spin-dependent) Fermi-Dirac function,
\begin{subequations}
\begin{align}
    f^\text{in}_{\kappa \uparrow}(\epsilon) &= f(\epsilon - \mu_0 - \mu^C_\kappa - \mu^{S}_{\kappa}/ 2) \\
    f^\text{in}_{\kappa \downarrow}(\epsilon) &= f(\epsilon - \mu_0 - \mu^C_\kappa + \mu^{S}_{\kappa}/ 2) \, .
\end{align}
\end{subequations}
For notational simplicity, we define the chemical potentials
\begin{subequations}
\label{eq:chemical_pots_spindep}
\begin{align}
    \mu_{\kappa \uparrow}   &= \mu_0 +  \mu^C_\kappa + \mu^{S}_{\kappa}/ 2\\
    \mu_{\kappa \downarrow} &= \mu_0 +  \mu^C_\kappa - \mu^{S}_{\kappa}/ 2 \, .
\end{align}
\end{subequations}
In the limit of small charge and spin accumulations compared to the Fermi level, it can be derived that 
\begin{align}
\label{eq:acc_in}
    \mu^C_\kappa + \frac{\mu^S_\kappa}{2} = \int d\epsilon 
    \left[ f_{\kappa \uparrow}^{\text{in}}(\epsilon) - f(\epsilon) \right] 
    \, .
\end{align}

In the elastic regime, the distribution function cannot generally be described as a Fermi-Dirac function.
The distribution function is instead given as a linear combination of Fermi-Dirac functions in the connected reservoirs~\cite{Brataas:PRB2022},
\begin{align}
    f^{el}_{s \kappa}(\epsilon) = \sum_l R_{s \kappa l} f(\epsilon - \mu_l) \, ,
\end{align}
where the index $l$ runs over the reservoirs, and  $R_{s \kappa l}$ is the lead and spin-dependent transport coefficient for reservoir $l$.
The transport coefficients satisfy 
\begin{align}
    \label{eq:sum_transport_unity}
    \sum_l R_{s \kappa l} = 1 \, .
\end{align}
In the elastic transport regime, it is advantageous to define the \textit{effective} charge and spin accumulations through
\begin{align}\label{eq:acc_el}
 \mu^C_\kappa + \frac{\mu^S_\kappa}{2} = \int d\epsilon 
    \left[ f_{\kappa \uparrow}^{\text{el}}(\epsilon) - f(\epsilon) \right] \, .
\end{align}
The elastic and inelastic transport regime results in different results for the fluctuations in the magnetization dynamics of the magnet.

Having specified the model for the system in consideration, we proceed by presenting the main results of the paper.

\section{Main results: Equations of motion}
\label{section:main_results}

This section presents the main results of our work.
Our primary result is the derivation of a 
Landau–Lifshitz–Gilbert–Slonczewski (LLGS) equation for the localized (normalized) spins $\vec m_i = \vec S_i/S_i$ in a general magnet coupled to normal metal reservoirs,
\begin{align}
\label{eq:eom_main}
    \partial_t \vec m_i &= \vec \tau^\text{b}_i + \vec \tau^\text{f}_i + \vec \tau^\text{sp}_i + \vec \tau_i^\text{stt}
\end{align}
valid for low-energy excitations when the equilibrium magnetization is parallel (antiparallel) to the $z$-axis. 
The bulk antiferromagnet torque $\vec \tau^\text{b}_i$ for a site $i = \{a, b\}$  arises from contributions of anisotropy, exchange coupling, and external fields, and reads
\begin{align}
\label{eq:bulk_torque_main}
    \vec \tau^\text{b}_i &= - \vec z \times \left(\hbar^{-1} E_i \vec m_{i} +  \gamma \mu_0 \vec H_{i}  \right) \, .
\end{align}
where $E_i$ is the energy of a local excitation and $\vec H_i$ is the applied field.
Hence, the bulk torque remains unaffected by the presence of normal metal reservoirs and the associated spin- and charge accumulations.

The spin transfer torque $\vec \tau^\text{stt}_i$ is induced by spin accumulation in the normal metals, and can be expressed as follows:
\begin{align}
\label{eq:stt_main}
    \vec \tau_i^\text{stt} =  \hbar^{-1} \sum_\kappa \left[\beta_{i \kappa}^I \vec z \times \vec \mu^S_\kappa - \beta_{i \kappa}^R \vec z \times (\vec z \times \vec  \mu^S_\kappa)\right] \, .
\end{align}
In \eq \eqref{eq:stt_main}, 
the superscripts ''$R$,$I$'' denote the real and imaginary part.
The site and lead-dependent coefficients $\beta_{i\kappa}$ are expressed in terms of the microscopic scattering matrix elements defined in \eq \eqref{W} evaluated at the Fermi energy:
\begin{align}
\label{eq:stt_coeffs_main}
    \beta_{i \kappa} = - \frac{2 \iu}{s_i}  \sum_{ n} W_{i \ud} ^{\kappa n \kappa n} \, ,
\end{align}
and can be calculated numerically for any particular system configuration.

The spin pumping torque $\vec \tau^\text{sp}_i$ contains contributions from both sublattices and is given by
\begin{equation}
\label{eq:tau_sp_main}
    \vec \tau^\text{sp}_i =  \sum_{j} \left[ \alpha_{ij}^R \vec z \times \partial_t \vec m_j 
    + \alpha_{ij}^I \vec z \times( \vec z \times \partial_t \vec m_j)
    \right]
    \, ,
\end{equation}
where $j$ runs over all sites and $\alpha_{ij}$ is expressed in the low-energy limit using the scattering matrix elements evaluated at the Fermi energy,
\begin{equation}
\label{eq:beta_main}
    \alpha_{ij} =  \frac{2 \pi}{\sqrt{s_i s_j}} \sum_{\kappa \lambda nm}  
    W_{i \du}^{\kappa n \lambda m} W_{j \ud  }^{\lambda m \kappa n} \, ,
\end{equation}
and $\alpha^{R(I)}$ denotes the real (imaginary) part of the matrix.
Using the Onsager reciprocal relations in Appendix \ref{app:onsager}, we find that the spin transfer torque and spin pumping are related in the case of the most relevant case of a single reservoir, 
\begin{align}
\label{eq:main_beta=alpha}
    \sum_{j} \alpha_{ij} = \beta_i \, .
\end{align}

Finally, the fluctuating torque $\vec \tau_i^\text{f}$ is expressed in terms of a fluctuating transverse field $\vec H^\text{f}_i$,
\begin{equation}
\label{eq:fluctuating_torque_main}
    \vec \tau^\text{f}_i =
    - \gamma \mu_0 \vec z  \times \vec H^\text{f}_i \, . 
\end{equation}
The fluctuating field exhibit interlattice and intralattice correlators $\langle H_{\mu i} H_{\nu j} \rangle$, where $\mu, \nu = \{x, y\}$: 
\begin{subequations}
\label{eq:fluctuations_main}
\begin{align}
     2 \sqrt{s_i s_j} \gamma^2 \mu_0^2 \langle H^\text{f}_{x i} H^\text{f}_{x j} \rangle &=   
    \Im \Sigma_{ij}^K
    + 4 \Im \tilde \Sigma_{\ud ij} 
    \\
    2 \sqrt{s_i s_j} \gamma^2 \mu_0^2 \langle H^\text{f}_{xi}  H^\text{f}_{y j}  \rangle &= - \Re \Sigma_{ij}^K  
    - 4 \Re \tilde \Sigma_{\ud ij}
    \\
    2 \sqrt{s_i s_j} \gamma^2  \mu_0^2 \langle H^\text{f}_{y i} H^\text{f}_{y j} \rangle &= \Im \Sigma_{ij }^K -
     4 \Im \tilde \Sigma_{\ud ij} \, ,
\end{align}
\end{subequations}
where the time arguments $t$ and $t'$ of the fields and the relative time argument $(t - t')$ of the self energies are omitted for simplicity. 
The self-energy $\Sigma$ is due to charge and longitudinal spin accumulations in the normal metals and is nonzero even in equilibrium. 
It is conveniently written as a product of a frequency-dependent quantity $\pi(\omega)$ and a site and scattering states dependent quantity $\sigma_{ij}$ \cite{Brataas:PRB2022}:
\begin{align}
\label{eq:main_sigma}
     \Sigma_{ij}^K(\omega) = \frac \iu \hbar \sum_{\kappa \lambda} \sigma_{ij \kappa \lambda}\pi_{\kappa \lambda}(\omega) \, ,
\end{align}
with 
\begin{subequations}
\begin{align}
\nonumber
    \pi_{\kappa \lambda  }(\omega) &= - 2 \int d\epsilon \left[2 n_{\uu \kappa}(\epsilon) n_{\dd \lambda}(\epsilon + \hbar \omega) \right. \\ 
    &\phantom{= - 2 \int d\epsilon [a}  \left.-  n_{\uu \kappa}(\epsilon) - n_{\dd \lambda}(\epsilon + \hbar \omega)
    \right] \\
    \sigma_{ij \kappa \lambda} &=  \frac{2 \pi}{\sqrt{s_i s_j}} \sum_{nm} W_{i \du}^{\kappa  n \lambda m } W_{j \ud}^{\lambda  m  \kappa n} \, .
\end{align}
\end{subequations}
Conversely, the self-energy matrices $\tilde \Sigma_{\ud}$ are due to transverse spin accumulation in the normal metals and, as a result, vanish in equilibrium.
Analogous to the decomposition in \eq \eqref{eq:main_sigma}, we write
\begin{equation}
    \tilde \Sigma_{ \ud ij}^K = - \frac \iu \hbar \sum_{\kappa \lambda} \tilde \sigma_{\ud ij \kappa \lambda} \tilde \pi_{\kappa \lambda}(\omega) \, ,
\end{equation}
where 
\begin{subequations}
\begin{align}
    \tilde \pi_{ \ud }(\omega) &= - 4 \int d\epsilon n_{\ud \kappa}(\epsilon) n_{\ud \lambda}(\epsilon + \hbar \omega)\\
    \tilde \sigma_{ \ud ij \kappa \lambda} &= - \frac{\pi}{\sqrt{s_i s_j}} \sum_{nm} W_{i \du}^{\kappa n \lambda  m} W_{j \du}^{\lambda m \kappa n}\, ,
\end{align}
\end{subequations}
The noise matrices $\pi(\omega)$ and $\tilde \pi(\omega)$ are similar to what was found in Ref.\ \onlinecite{Brataas:PRB2022}, and are calculated in the equilibrium, elastic, and inelastic transport regime in Sec. \ref{section:second_order}.
Crucially, the shot noise differs on various sites, due to the site-dependence of $\sigma$ and $\tilde \sigma$.
At equilibrium, the fluctuation-dissipation theorem holds, e.g. 
\begin{align}
    2 s_i \gamma^2 \mu_0^2 \langle H^\text{f}_{\mu i} H^\text{f}_{\nu i} \rangle =  \delta_{\mu \nu} \alpha_{ii} 4 k_B T \xi \left(\frac{\hbar \omega}{2 k_B T} \right),
\end{align}
where $\xi(x) = x \coth x$.

In the next section, we discuss the Keldysh action of the model presented in this section and derive an effective action by integrating out the fermionic degrees of freedom.

\section{Keldysh theory and effective action}
\label{section:effective_action}
In this section, we derive the semiclassical spin dynamics by using an out-of-equilibrium path integral formalism~\cite{kamenev_field_2011}.
We introduce the closed contour action $S$ and the partition function $Z$,
\begin{align}
\label{eq:Z}
    Z = \int D[\bar{a} a \bar{b} b \bar{c}_\uparrow c_{\uparrow} \bar{c}_\downarrow c_\downarrow] \e^{  \iu  S / \hbar} \, .
\end{align}
The action $S$ consists of contributions from the localized magnetic excitations $a$ and $b$, and the spin-up $c_\uparrow$ and spin-down electrons $c_\downarrow$ from the scattering states.
We will integrate out the fermion operators and get an effective action for the magnetic excitations $a$ and $b$, which includes effective transverse and longitudinal fields that arise from the charge and spin accumulations in the normal metals.

We follow Ref.\ \onlinecite{kamenev_field_2011} and replace the fields in \eq \eqref{eq:Z} with ''$\pm$'' fields residing on the forward and backward part of the Schwinger-Keldysh contour. The action in the $\pm$ basis is given in Appendix~\ref{appendix:pm}.
These fields are not independent of each other and can be Keldysh rotated into a new basis that takes into account the coupling between them. The rotated fields have the advantage of suggesting a transparent physical interpretation, corresponding to the semiclassical equations and quantum corrections.

\subsection{Keldysh action}

For magnons, the classical ($cl$) and quantum ($q$) fields are defined linear combinations of the $\pm$-fields, as described in detail in Appendix \ref{appendix:pm}.
In Keldysh space, it is convenient to also introduce the matrices
\begin{align}
    \gamma^q = \begin{pmatrix}
    0 & 1\\
    1 & 0
    \end{pmatrix}
    &&
    \gamma^{cl} = \begin{pmatrix}
    1 & 0 \\
    0 & 1
    \end{pmatrix} \,.
\end{align}
The Keldysh rotated magnon action becomes 
\begin{align}
    \nonumber
    S_m
    =& 
     \sum_{ a t} \bar  a_a^q( \iu \hbar \partial_t - E^A_{a})  a_a^{cl}  
     +
     \sum_{ b t} \bar b_b^q ( \iu \hbar \partial_t - E^B_{b})    b_b^{cl} 
     \\
    \nonumber
    &
    + \sum_{ a t}   a_a^q( \iu \hbar \partial_t - E^A_{a})  \bar a_a^{cl}  
    +
     \sum_{bt}  b_b^q ( \iu \hbar \partial_t - E^B_{b})   \bar b_b^{cl} 
    \\ \nonumber
    &
    -
    2 \sum_{aa' t} \sqrt{s_a s_{a'}} J_{aa'} \left[\bar{a}^{q}_a  a_{a'}^{cl} + h.c. \right]
    \\
    &
    - 
    2 \sum_{bb't} \sqrt{s_b s_{b'}} J_{bb'}
    \left[ \bar b_b^{q}  b_{b'}^{cl} + h.c \right]
    \nonumber
    \\
    \nonumber
    &
    -
    2 \sum_{ab t} \sqrt{s_a s_{b}} J_{ab}   \left[\bar a_a^q  \bar b_b^{cl} + \bar a_a^{cl}  \bar b_b^{q} + h.c. \right]
    \\ 
    \nonumber
    &
    -
    \gamma \mu_0 \hbar \sum_{a t} \sqrt{s_a} \left[ H^A_{a-}  a_a^q + H^A_{a+} \bar a_a^q \right]
    \\    &
    -
    \gamma \mu_0 \hbar \sum_{b t} \sqrt{s_b}   \left[ H^B_{b-} \bar b_b^q + H^B_{b+} b_b^q  \right] \, ,
    \label{eq:Smrot}
\end{align}
where we wrote the time integral as a sum for compact notation.
In \eq \eqref{eq:Smrot}, $h.c.$ denotes the hermitian conjugate of the previous term.
The fermion action becomes
\begin{equation}
    S_e + S_0 =
   \sum_{s t}
     \bar C_s \gamma^{cl} (i \hbar \partial_t - \epsilon) C_{s}
     \, ,
    \label{SeRot}
\end{equation}
where we introduced vector notation for the $1/2$-fields, $\bar C_{s} = (\bar c_{s \alpha}^{1}, \bar c_{s \alpha}^2)$, and $\epsilon$ is a diagonal matrix containing the single-particle energies of the electrons. 
The Keldysh rotated first-order electron-magnon interaction is
\begin{widetext}
\begin{align}
        S_1 =&  -\sum_{\substack{a t \\ \alpha \beta} }  
        \frac{1}{\sqrt{s_a}}  
        \left[
        {a}_a^{cl} W_{a \downarrow \uparrow}^{\alpha \beta}   \bar{C}_{\downarrow \alpha}\gamma^{cl}   {C}_{\uparrow \beta}  + 
        {a}_a^{q} W_{a \downarrow \uparrow}^{\alpha \beta} \bar{C}_{\downarrow \alpha}\gamma^q {C}_{\uparrow \beta} 
        + h.c. \right]
        \nonumber
        \\
        &  - \sum_{\substack{b t \\ \alpha \beta} }  \frac{1}{\sqrt{s_b}}  
        \left[
        {\bar b}_a^{cl} W_{b \downarrow \uparrow}^{\alpha \beta}   \bar{C}_{\downarrow \alpha} \gamma^{cl}   {C}_{\uparrow \beta} +
        {\bar b}_b^{q} W_{b \downarrow \uparrow}^{\alpha \beta} \bar{C}_{\downarrow \alpha} \gamma^q   {C}_{\uparrow \beta}
        + h.c. \right]\,,
        \label{S1Rot}
\end{align}
and, finally, the second-order term reads
\begin{align}
    \nonumber
    S_2 =& 
    \sum_{\substack{ a t \\ \alpha \beta}}
    \frac{1}{2 s_a} 
    \left[
    W_{a \uparrow \uparrow}^{\alpha \beta}  \left( \bar A_a \gamma^{cl} A_a  
     \bar 
     C_{\uparrow \alpha}\gamma^{cl} {C}_{\uparrow \beta} 
    +  \bar A_a \gamma^q A_a \bar C_{\uparrow \alpha}\gamma^q {C}_{\uparrow \beta}
    \right) 
    -
    W_{a\downarrow \downarrow}^{\alpha \beta}  \left( \bar A_a \gamma^{cl} A_a  
     \bar C_{\downarrow \alpha}\gamma^{cl} {C}_{\downarrow \beta} 
    +  \bar A_a \gamma^q A_a \bar C_{\downarrow \alpha}\gamma^q {C}_{\downarrow \beta}
    \right) 
    \right]
    \\ 
    & 
    -\sum_{\substack{ b t \\ \alpha \beta}}
    \frac{1}{2 s_b}
    \left[
    W_{b \uparrow \uparrow}^{\alpha \beta}  \left( \bar B_b \gamma^{cl} B_b 
     \bar C_{\uparrow \alpha}\gamma^{cl} {C}_{\uparrow \beta} 
    +  \bar B_b \gamma^q B_b \bar C_{\uparrow \alpha}\gamma^q {C}_{\uparrow \beta}
    \right) 
    -
    W_{b \downarrow \downarrow}^{\alpha \beta}  \left( \bar B_b \gamma^{cl} B_b  
     \bar C_{\downarrow \alpha}\gamma^{cl} {C}_{\downarrow \beta} 
    +  \bar B_b \gamma^q B_b \bar C_{\downarrow \alpha}\gamma^q {C}_{\downarrow \beta}
    \right)
    \right]  
    \, ,
\end{align}
\end{widetext}
where the magnon $q/cl$ operators are consolidated in vectors $\bar A_a$ and $\bar B_b$.
The Keldysh rotated action proves to be well-suited for the computation of an effective magnon action, a topic we delve into in the following section.

\subsection{Integrating out the fermionic degrees of freedom}
For the itinerant electrons in the normal metal and antiferromagnet, the total effective electron action is $S_{e, \text{tot}} = S_e + S_{em}$, and can be expressed as
\begin{equation}
    \label{S_eeff}
     S_{e, \text{tot}} =
    \sum_{ss'  tt'} 
    \bar C_{s, t}
    G^{-1}_{s s',tt'}
    C_{s', t'}\, ,
\end{equation}
where the interacting Green function $G$ is given in terms of the noninteracting Green function $G_0$ and interaction terms as
\begin{equation}
\label{Gint}
    G^{-1} =  G_0^{-1} +\tilde W_1 + \tilde W_2 \, .
\end{equation}
Here, $\tilde W_1$ contains the first-order magnon operators on both sublattices, 
\begin{equation}
    \tilde W_1 = \delta(t - t') \left[W_1^A + W_1^B\right] \, ,
\end{equation}
where $W_1^A$ and $W_1^B$ are spin flip operators:
\begin{subequations}
\begin{align}
    W_1^A &= -  \sum_{x a} \frac{1}{\sqrt{s_a}} \gamma^x  \bigl[W_{a \uparrow \downarrow} \bar a^x_a \sigma_+ 
    + W_{a \downarrow \uparrow } a^x_a \sigma_- \bigr]
    \\
    W_1^B&= -  \sum_{xb} \frac{1}{\sqrt{s_b}} \gamma^x  \bigl [W_{b\uparrow \downarrow} b^x_b \sigma_+ 
    + W_{ b\downarrow \uparrow}\bar b^x_b \sigma_-\bigr].
\end{align}
\label{W1}
\end{subequations}
In \eq \eqref{W1}, the variable $x = \left\{cl, q\right\}$ represents a Keldysh space index, and $\sigma_\pm$ are the usual raising and lowering Pauli matrices.
Similarly, $\tilde W_2$ contains the magnon operators to quadratic order for both sublattices, 
\begin{equation}
    \tilde W_2 = \delta(t - t') \left[W_2^A - W_2^B\right] \, ,
\end{equation}
with $W_2^A$ and $W_2^B$ given by
\begin{subequations}
\begin{align}
    W_2^A &= \sum_{a x y} \frac{1}{2 s_a} \bar a^x_a \gamma^x a^{y}_a \gamma^{y} 
    \begin{pmatrix}
        W_{a\uu} & 0
        \\
        0 & -W_{a\dd}
    \end{pmatrix}
    \\
    W_2^B &= \sum_{b x y}\frac{1}{2 s_b}  \bar b^x_b \gamma^x b^{y}_b \gamma^{y} 
    \begin{pmatrix}
        W_{b\uu} & 0
        \\
        0 & -W_{b\dd}
    \end{pmatrix} \, ,
\end{align}
\end{subequations}
where the spin structure is explicitly written out as a matrix.
The matrices $W_1^{A(B)}$ and $W_2^{A(B)}$ have a structure in the scattering states space from $W_{a(b)}$, spin space from the Pauli matrices, and Keldysh space from $\gamma^x$. 
The inverse free electron Green function $G_0^{-1}$ from \eq \eqref{Gint} has the conventional causality structure in Keldysh space, with a retarded ($R$), advanced ($A$), and Keldysh ($K$) component:
\begin{equation}
\label{G_inv}
    G^{-1}_{0} = 
    \begin{pmatrix}
    [G^{R}_{0}]^{-1} & [G^{K}_{0}]^{-1}\\
    0 & [G^{A}_{0}]^{-1}
    \end{pmatrix}\, ,
    \end{equation}
and has equilibrium components that are diagonal in both spin space and in the scattering states space, 
\begin{equation}
    [G_0^{-1}]^{R(A)}_{\alpha \beta, ss'} = 
    \delta_{\alpha \beta} \delta_{ss'} \delta(t - t') 
    ( \iu \hbar \partial_t - \epsilon_\alpha \pm i \delta) 
    \, ,
\end{equation}
where the upper sign corresponds to the retarded component, while the lower sign is applicable to the advanced component.
The Keldysh component includes information about the distribution function, and will be discussed below, when we Fourier transform the Green functions.

From the effective electron action in \eq \eqref{S_eeff}, it is evident that the partition function of $S_{e, \text{tot}}$ takes on a Gaussian form with respect to the fermionic operators. Hence, the fermionic integral in the partition function can be evaluated exactly, with an inconsequential proportionality constant being disregarded:
\begin{equation}
\label{eq:exact_fermion_int}
    \int D[C]
    \e^{ \iu  S_{e, \text{tot}}/\hbar} = \e^{\tr{\ln \left[1 + G_0 \tilde W_1 + G_0 \tilde W_2 \right]}}\, .
\end{equation}
In \eq \eqref{eq:exact_fermion_int}, we have used the short-hand notation for the functional integral measure of all fermionic states, $D[C] = D[ \bar{C}_\uparrow C_{\uparrow} \bar{C}_\downarrow C_\downarrow] $. We have absorbed a normalization constant into the functional integral measure for simplicity.
We note that as a consequence of the continuity of the time coordinate and scattering states energy that we are employing, the unit matrix is a delta function in time and energy, $1 \equiv \delta(t - t') \delta(\epsilon_\alpha - \epsilon_\beta)$, and thus quantities inside the logarithm carries dimension \si{\joule^{-1} \second^{-1}}. The trace, on the other hand, is an integral operator with unit \si{\joule \second}. As long as one interprets the logarithm in terms of its Taylor expansion, this does not lead to any problems, as the exponent of \eq \eqref{eq:exact_fermion_int} becomes dimensionless for all terms in the expansion.
The exponent is interpreted as an additional contribution to the magnon action, 
\begin{equation}
\frac \iu \hbar  S_\text{eff} =  \tr{\ln [1 + G_0 \tilde W_1 + G_0 \tilde W_2 ]} \, .
\end{equation}
The way forward is to treat this interaction as a perturbation, expanding the logarithm in first and second-order contributions and disregarding higher-order terms,
\begin{align}
    \nonumber
     S_\text{eff} \approx& - \iu \hbar \tr{ G_0 \tilde W_1 }         - \iu \hbar \tr{ G_0 \tilde W_2} 
    \\
    &
    + \frac{\iu \hbar}{2} \tr{G_0 \tilde W_1 G_0 \tilde W_1 } \, . 
    \label{Seff_expansion}
\end{align}
To evaluate the trace in these terms, it is convenient to Fourier transform all quantities from the time domain to the energy domain. This diagonalizes the noninteracting Green functions, making calculations much more straightforward. 

\subsection{Fourier representation}
The paper employs the Fourier transform convention defined in Appendix \ref{appendix:fourier}.
In Fourier space, the fermion equilibrium Green function components are particularly simple:
\begin{align}
    [G_0]^{R(A)}_{\alpha \beta, ss'}(\omega) = \delta_{\alpha \beta} \delta_{ss'}  ( \hbar \omega - \epsilon \pm \iu \delta)^{-1} \, ,
\end{align}
where $\delta>0$ is an infinitesimal quantity ensuring convergence.
The Keldysh component accounts for non-equilibrium phenomena though the spin-dependent distribution $n_{ss' \alpha}$ defined in \eq \eqref{n_ssa},
\begin{equation}
    [G_0]^{K}_{\alpha \beta, ss'}(\omega) = - 2 \pi \iu \delta_{\alpha \beta} \delta(\hbar \omega - \epsilon_\alpha) \left[ \delta_{ss'} - 2 n_{ss' \alpha} \right] .
\end{equation}
The Keldysh component has off-diagonal terms in spin space if the distribution function $n_{ss' \alpha}$ has off-diagonal elements, i.e. if there is a transverse spin accumulation in the normal metals.

\section{Non-equilibrium Spin Dynamics}
\label{section:spindynamics}

Having derived the effective action as expressed in \eq \eqref{Seff_expansion}, we proceed by evaluating the traces and delving into the resultant terms. The discussion unveils effective longitudinal and transverse fields, which we ascribe to spin transfer torque and spin pumping originating from the normal metal reservoirs.

\subsection{First-order contribution}
Evaluating the trace in the first order term in \eq \eqref{Seff_expansion} corresponds to summing over the diagonal elements in spin space and Keldysh space, integrating over both time variables, and summing over the space of scattering states, we find 
\begin{align}
    \nonumber
    - \iu \hbar \tr{G_0 \tilde  W_1} =& - \sum_{a \alpha }
    \frac{2}{\sqrt{s_a}}
    W_{a \uparrow \downarrow}^{\alpha \alpha} n_{\downarrow \uparrow \alpha} \int dt \bar a^{q}_a(t)
    \\
    \nonumber
    & - \sum_{a\alpha }
    \frac{2}{\sqrt{s_a}}
    W_{a \downarrow \uparrow}^{ \alpha \alpha} n_{\uparrow \downarrow \alpha}  \int dt a^{q}_a(t) 
    \\
    \nonumber
    & - \sum_{b\alpha }
    \frac{2}{\sqrt{s_a}}
    W_{b  \uparrow \downarrow}^{ \alpha \alpha} n_{\downarrow \uparrow \alpha}  \int dt b^{q}_b( t) 
    \\
    & - \sum_{b\alpha }
    \frac{2}{\sqrt{s_a}}
    W_{b \downarrow \uparrow}^{ \alpha \alpha} n_{\uparrow \downarrow \alpha}  \int dt \bar b^{q}_b(t)\, .
    \label{S1_final}
\end{align}
Here, we have used the general Green function identity $G^R(t, t) + G^A(t,t) = 0$ \cite{kamenev_field_2011}, and written the time integration explicitly.
Comparing the first-order contribution in \eq \eqref{S1_final} with the magnon action in \eq \eqref{eq:Smrot}, we observe that the first-order effect of the spin accumulation in the normal metal is equivalent to an effective deterministic transverse magnetic field $\vec H_i^\text{stt}$, which act on a localized spin at site $i = \{a, b\}$ in the antiferromagnet.
The ''stt'' superscript indicates that this field will take the form of a spin transfer torque, which will be elaborated on below.
The magnitudes of these effective transverse fields are given by
\begin{subequations}
\label{eq:Hpms}
\begin{align}
    \gamma \mu_0 H_{i-}^\text{stt} &= \frac{2}{s_i \hbar} \sum_\alpha W_{i \downarrow \uparrow}^{\alpha \alpha} n_{ \uparrow \downarrow \alpha}
    \\
    \gamma \mu_0 H_{i +}^\text{stt} &= \frac{2}{s_i \hbar} 
    \sum_\alpha W_{i \uparrow \downarrow}^{\alpha \alpha}n_{\downarrow \uparrow \alpha}  \, ,
\end{align}
\end{subequations}
which implies that the Cartesian components read
\begin{subequations}
    \label{Hxy_firstorder}
    \begin{align}
        \gamma \mu_0 H_{i x}^\text{stt} &= \frac{2}{s_i \hbar} \sum_\alpha \Re \left[W_{i \ud}^{\alpha \alpha} n_{\du \alpha}
        \right]
        \\
        \gamma \mu_0 H_{i y}^\text{stt} &= \frac{2}{s_i \hbar} \sum_\alpha \Im \left[W_{i \ud}^{\alpha \alpha} n_{\du \alpha}
        \right] \, .
    \end{align}
\end{subequations}
Recalling that the spin accumulation is given by \eq \eqref{eq:acc_el} and \eq \eqref{eq:acc_in}, we write the effective fields from \eq \eqref{Hxy_firstorder} in the conventional spin transfer torque form:
\begin{align}
\label{eq:transverse_fields}
    \gamma \mu_0 \vec H^\text{stt}_i &= \frac{1}{\hbar} \sum_\kappa
    \left[
        \beta_{i\kappa}^R  \vec z \times \vec \mu^S_\kappa 
        + 
        \beta_{i \kappa}^I \vec z \times (\vec z \times \vec \mu^S_\kappa)
    \right] \, ,
\end{align}
where the appearance of $\vec z$ is a consequence of our theory being restricted to small deviations for the equilibrium magnetization $\pm \vec z$.
This results in the spin transfer torque given in \eq \eqref{eq:stt_main}.
In \eq \eqref{eq:transverse_fields}, the superscripts ''$R$'' and ''$I$'' denote the real and imaginary parts and the lead- and site-dependent constants $\beta_{i \kappa}$ have been introduced as sums over the transverse modes of the scattering matrix elements,
\begin{align}
    \label{eq:alpha}
    \beta_{i \kappa} =  -\frac{2 \iu }{s_i} \sum_{ n} W_{i \ud} ^{\kappa n \kappa n} 
    \, ,
\end{align}
and where we have assumed that the transverse spin distribution functions $n_{\ud}$ and $n_{\du}$ are only significant close to the Fermi surface, such that the scattering states matrix elements are well approximated by their value at the Fermi surface.
The expression for the spin transfer field in \eq \eqref{eq:transverse_fields} is valid in both the elastic and inelastic regime, and vanishes in equilibrium.
We note that the coefficient $\beta_{i \kappa}$, for $i = \{a, b\}$, depends not only on the potential at lattice site $i$ but also indirectly of all lattice sites on both sublattices through the scattering states.

To the lowest order, the sublattice magnetizations are parallel and antiparallel to the $z$-axis, $\vec m_A \approx \vec z$ and $\vec m_B \approx - \vec z$. 
Thus, to the lowest order in the magnon operators, the expressions for the transverse fields are ambiguous, and we can write the transverse field in \eq \eqref{eq:transverse_fields} in terms of $\vec m_A$ or $\vec m_B$.
To the lowest order in the magnon operators, the Keldysh technique cannot be used to identify which sublattice the transverse fields in \eq \eqref{S1_final} originate from.

\subsection{Second order contribution}
\label{section:second_order}
The second order contribution in \eq \eqref{Seff_expansion} has contributions 
from $\tilde W_2$,
\begin{equation}
    S_{21} = - \iu \hbar \tr{G_0 \tilde W_2} \, ,
\end{equation}
as well as a contribution from $\tilde W_1$,
\begin{equation}
    S_{22} = \frac{\iu \hbar}{2} \tr{G_0 \tilde W_1 G_0 \tilde W_1 } \, .
\end{equation}
Proceeding in a manner analogous to the treatment of the first-order term, the trace in $S_{21}$ is evaluated: 
\begin{widetext}
\begin{align}
        S_{21} =& - \sum_{a \alpha} 
        \frac{\pi}{s_a}
        \bigl[W_{a \uu}^{\alpha \alpha}( 1 - 2 n_{\uu \alpha}) - W_{a \dd}^{\alpha \alpha} (1 - 2 n_{\dd \alpha})\bigr]
        \int dt \bar A_a(t) \gamma^q A_a(t) 
        \nonumber
        \\
        &
        + \sum_{b \alpha} \frac{\pi}{s_b} \bigl[W_{b \uu}^{\alpha \alpha}( 1 - 2 n_{\uu \alpha}) - W_{b \dd}^{\alpha \alpha} (1 - 2 n_{\dd \alpha})\bigr]
        \int dt \bar B_b(t) \gamma^q B_b(t) \, .
        \label{eq:S21}
\end{align}
From \eq \eqref{eq:Smrot}, it is apparent that the second-order terms in $S_{21}$ are equivalent with a longitudinal magnetic field, with magnitude 
\begin{align}
    \gamma \mu_0 H_{i z}^{A21} &= -\frac{\pi}{\hbar s_i } \sum_{\alpha} \left[W_{i \uu}^{\alpha \alpha}( 1 - 2 n_{\uu \alpha}) - W_{i \dd}^{\alpha \alpha} (1 - 2 n_{\dd \alpha})\right] \, ,
\end{align}
which, in this reference frame, renormalizes the energies of localized magnon excitations. 
However, such longitudinal magnetic fields should not affect the spin dynamics since they, in the instantaneous reference field, correspond to contributions to the total free energy proportional to ${\bf S}_i^2$.

The final contribution $S_{22}$ to the effective action contains inter-lattice and intra-lattice terms and can be written compactly by introducing a field $d_i = \{a_a, \bar b_b \}$ and summing over the two field components, i.e. $\sum_i d_i= \sum_{a} a_a + \sum_b \bar b_b$:
\begin{align}
    \nonumber
    S_{22} =&  \int dt dt' \sum_{x x' ij} 
    \frac{\iu \hbar}{2\sqrt{s_i s_j}}
     \tr{G_{0, \ud}(t', t)  W_{i \du} \gamma^{x} G_{0, \ud}(t, t') \gamma^{x'} W_{j \du}}  d^x_i(t)  d^{x'}_{j}(t')
     \\
     \nonumber
     &
     +
     \int dt dt' \sum_{x x'ij} 
     \frac{\iu \hbar}{2\sqrt{s_i s_j}}
     \tr{G_{0, \du}(t', t)  W_{i \ud} \gamma^{x} G_{0, \du}(t, t') \gamma^{x'} W_{j \ud}  } \bar d^x_i(t) \bar d^{x'}_{j}(t')
     \\
     \nonumber
     & 
     +
    \int dt dt' \sum_{x x'ij} 
     \frac{\iu \hbar}{2\sqrt{s_i s_j}}
     \tr{G_{0, \dd}(t', t)  W_{i \du} \gamma^{x} G_{0, \uu}(t, t') \gamma^{x'} W_{j \ud}    } d^x_i(t)  \bar d^{x'}_{j}(t')
     \\
     &
     +
     \int dt dt' \sum_{x x'ij}
      \frac{\iu \hbar}{2\sqrt{s_i s_j}}
     \tr{G_{0, \uu}(t', t)  W_{i \ud} \gamma^{x} G_{0, \dd}(t, t') \gamma^{x'} W_{j \du}   } \bar d^x_i(t)  d^{x'}_{j}(t') \, ,
    \label{S22dd}
\end{align}
where the trace is taken only over the $2\times2$ Keldysh space and the space of scattering states $\alpha$. 
The interlattice terms, i.e. $d = d'$, are discussed in Ref.~\onlinecite{Brataas:PRB2022} for a macrospin ferromagnet. Here, we summarize this discussion and highlight the addition of the inter-lattice terms not present in the macrospin ferromagnet.

Evaluating the trace in the first and second line of \eq \eqref{S22dd}, we note that only the Keldysh component has off-diagonal elements in spin space, and find a contribution only from $x = x' = q$,
\begin{subequations}
\label{S22qq}
\begin{align}
    \tilde S_{22}^{qq} &= \hbar  \int dt dt' \sum_{ ij } 
    \left[ d^q_i(t) \tilde \Sigma_{\ud ij}^K(t, t') d^{q}_{j}(t') \right]
    \\
    \tilde S_{22}^{\bar q \bar q} &= \hbar \int dt dt' \sum_{ ij } 
    \left[\bar d^q_i(t) \tilde \Sigma_{\du ij}^K(t, t') \bar d^{q}_{j}(t')
           \right] \, ,
\end{align}
\end{subequations}
where the self-energies are
\begin{subequations}
\label{eq:tilde_sigmas_definitions}
\begin{align}
    \tilde \Sigma_{\ud ij}^K(t - t') &= -\frac{2 \iu }{ \hbar^2 \sqrt{s_i s_j}} \sum_{\alpha \beta } n_{\ud \alpha} n_{\ud \beta} W_{i \du}^{\alpha \beta} W_{j \du  }^{\beta \alpha} \e^{ \iu   (\epsilon_\alpha - \epsilon_\beta) ( t - t')/\hbar}
    \\
    \tilde \Sigma_{\du ij}^K(t - t') &= -\frac{2 \iu }{ \hbar^2 \sqrt{s_i s_j}} \sum_{\alpha \beta } n_{\du \alpha} n_{\du \beta} W_{i \ud}^{\alpha \beta} W_{j \ud  }^{\beta \alpha} \e^{ \iu  (\epsilon_\alpha - \epsilon_\beta) ( t - t') / \hbar } \, .
\end{align}
\end{subequations}
The reasoning behind identifying this self-energy as a Keldysh component is that it couples the quantum components of the fields, see \eq \eqref{S22qq}.
The terms in \eq \eqref{S22qq} do not have a direct analog in the magnon action in \eq \eqref{eq:Smrot}, and interpreting these will be the subject of Sec.~\ref{section:fluctuating}.
The self-energies in \eq \eqref{eq:tilde_sigmas_definitions} are invariant under a joint time and lattice site reversal, i.e. $\tilde \Sigma_{ij}(t - t') = \tilde \Sigma_{ji}(t' - t)$.
Moreover, due to the properties $n_{\ud} = n_{\du}^*$ and $W_{i \ud}^{\alpha \beta } = [W_{i \du}^{\beta \alpha}]^*$, we see that the self-energies are related by $\tilde \Sigma_{\ud ij}^K(t - t') = -[\tilde \Sigma_{\du ij}^K(t - t')]^*$, which will be important later.

Disregarding terms of the order $k_B T / \epsilon_F$ and $\mu_s/\epsilon_F$ \cite{Brataas:PRB2022}, we find that the Fourier-transformed self-energy becomes
\begin{equation}
    \tilde \Sigma_{\ud ij }^K(\omega) = - \frac \iu \hbar \sum_{\kappa \lambda} \tilde \sigma_{\ud ij \kappa \lambda} \tilde \pi_{\kappa \lambda}(\omega) \, ,
\end{equation}
where 
\begin{align}
    \tilde \pi_{\kappa \lambda}(\omega) = - 4 \int d\epsilon n_{\ud \kappa}(\epsilon) n_{\ud \lambda}(\epsilon + \omega)&&
    \tilde \sigma_{ \ud ij \kappa \lambda} = - \frac{\pi}{\sqrt{s_i s_j}} \sum_{nm} W_{i \du}^{\kappa n \lambda m} W_{j \du}^{\lambda m \kappa n}\, ,
\end{align}
and where the matrix elements $W$ are evaluated at the Fermi surface.
This is a straightforward generalization of the macrospin ferromagnet case, with the addition of shot-noise contributions from inter-lattice and intra-lattice interactions between different lattice sites.
We can evaluate the quantity $\tilde \pi_{\kappa \lambda \ud }(\omega)$ by using \eq \eqref{n_ssa}:
\begin{align}
    \tilde \pi_{ \ud \kappa \lambda}(\omega) = - u_{\kappa -} u_{\lambda -}\int d\epsilon \left[ f_{ \uparrow \kappa}(\epsilon) - f_{\downarrow \kappa }(\epsilon) \right]
    \left[ f_{ \uparrow \lambda}(\epsilon + \hbar \omega) - f_{ \downarrow \lambda}(\epsilon + \hbar \omega) \right]
\end{align}
where we introduced the conventional ''lowering'' vector $u_{-} = u_x - \iu u_y$. This can be computed in equilibrium, elastic, and inelastic scattering cases, and results exactly similar to those in Ref.\ \onlinecite{Brataas:PRB2022}.

We now turn our attention to the third and fourth lines of the second-order action in \eq \eqref{S22dd}. The contributions from the two lines are equal, which is evident from interchanging summation indices and rearranging terms. 
Their total contribution to the action $S_{22}$ can be split into contributions $ S_{22}^{\bar qq}$, $S_{22}^{\bar q cl}$, and $S_{22}^{\bar{cl} q}$.
The contribution $S^{cl cl}$ vanishes, due to the quantity $G^{R}(t' - t) G^{R}(t- t')$ being nonzero only for $t = t'$, which has measure zero, and similarly for $G^A$. This ensures that the action satisfies the general requirement $S[\phi^{cl}, \phi^q = 0] = 0$~\cite{kamenev_field_2011}.
Introducing, for notational convenience, the vector $\bar D_i = \begin{pmatrix} \bar d^{cl} & \bar d^q \end{pmatrix}$, we find
\begin{align}
\label{eq:S22_not_tilde}
    S_{22}^{\bar qq}+ S_{22}^{\bar q cl} + S_{22}^{\bar{cl} q} &=  \hbar \int dt dt' \sum_{ ij } \bar D_i(t)
    \hat \Sigma_{ij}(t - t')  D_{j}(t') \, , 
\end{align}
where the self-energy matrix has structure in Keldysh space and in the sublattice space,
\begin{equation}
    \hat \Sigma_{ij}(t - t') = \begin{pmatrix}
        0 & \Sigma^A(t - t') \\
        \Sigma^R(t - t')  & \Sigma^K(t - t') 
    \end{pmatrix}_{ij} \, ,
\end{equation}
and its components are given by
\begin{subequations}
\begin{align}
    \label{eq:SigmaK}
     \Sigma_{ ij}^K(t - t') &= \frac{2 \iu }{\sqrt{s_i s_j} \hbar^2} \sum_{\alpha \beta } \left(n_{\uu \alpha} + n_{\dd \beta} - 2 n_{\uu \alpha} n_{\dd \beta}
     \right)
     W_{i \du}^{\alpha \beta} W_{j \ud  }^{\beta \alpha} \e^{\iu  (\epsilon_\alpha - \epsilon_\beta) ( t - t')/\hbar }
     \\
    \label{eq:SigmaR}
     \Sigma_{ ij}^R(t - t') &= \frac{2 \iu }{\sqrt{s_i s_j} \hbar^2} \theta( t - t') \sum_{\alpha \beta } \left(n_{\uu \alpha} - n_{\dd \beta}
     \right)
     W_{i \du}^{\alpha \beta} W_{j\ud  }^{\beta \alpha} \e^{\iu  (\epsilon_\alpha - \epsilon_\beta) ( t - t')/\hbar }
     \\
     \label{eq:SigmaA}
     \Sigma_{ ij}^A(t' - t) &= -\frac{2 \iu }{\sqrt{s_i s_j} \hbar^2} \theta( t - t') \sum_{\alpha \beta } \left(n_{\uu \alpha} - n_{\dd \beta}
     \right)
     W_{i \du}^{\alpha \beta} W_{j \ud  }^{\beta \alpha} \e^{\iu  (\epsilon_\alpha - \epsilon_\beta) ( t - t')/\hbar }\, .
\end{align}
\end{subequations}
The Keldysh component of this self-energy has the symmetry $ [\Sigma^K_{ij}(t- t')
]^* = -\Sigma^K_{ji}(t' - t)$. 
Imperatively, as a consequence of this symmetry, the quantities
\begin{align}
    \Sigma^K_{ ij}(t- t')-\Sigma^K_{ji}(t' - t) &= 2 \Re\left[ \Sigma^K_{ij}(t- t') \right]\\
    \iu \Sigma^K_{ ij}(t- t')  + \iu \Sigma^K_{ji}(t' - t) &= -2 \Im \left[ \Sigma^K_{ij}(t- t') \right] \, ,
\end{align}
are real numbers, which will be important in the next section.
We proceed by a similar analysis to what was done with $\tilde \Sigma$, writing it in terms of a shot-noise matrix.
We assume that the matrices $W$ can be approximated by their value on the Fermi surface, and write
\begin{align}
     \Sigma_{ij}^K(\omega) = \frac \iu \hbar \sum_{\kappa \lambda}  \sigma_{ij \kappa \lambda} \pi_{\kappa \lambda}(\omega) \, ,
\end{align}
where we introduced the matrices
\begin{align}
    \pi_{\kappa \lambda  }(\omega) = - 2 \int d\epsilon \left[2 n_{\uu \kappa}(\epsilon) n_{\dd \lambda}(\epsilon + \hbar \omega) -  n_{\uu \kappa}(\epsilon) - n_{\dd \lambda}(\epsilon + \hbar \omega)
    \right]&&
    \sigma_{ij \kappa \lambda} =  \frac{2 \pi}{\sqrt{s_i s_j}} \sum_{nm} W_{i \du}^{\kappa n \lambda m} W_{j \ud}^{\lambda m \kappa n}\, .
\end{align}
The matrix $\pi(\omega)$ can also be evaluated in equilibrium, and for elastic and inelastic scattering, and the results are again exactly similar to those in Ref.\ \onlinecite{Brataas:PRB2022}.
\end{widetext}
where $\xi(x) = x \coth x$ is an asymptotically linear function for high $x$ and $p_{ss'\kappa} = (1 - u_{z \kappa})/2 + u_{z \kappa} \delta_{ss}$ is a projection factor introduced for notational convenience. 

Comparing with the magnetic action in \eq \eqref{eq:Smrot}, we notice that the terms with the retarded and advanced self-energies are equivalent with longitudinal fields, which we in the following will show consists of dissipative Gilbert-like terms and non-dissipative field-like terms.
Fourier transforming and applying the identity \eqref{eq:ap:heaviside}, the retarded and advanced self-energies from \eq \eqref{eq:SigmaR} and \eq \eqref{eq:SigmaA} become
\begin{align}
\label{eq:SigmaRomega}
    \Sigma_{ij}^{R, A}&=  \frac{-2}{\sqrt{s_i s_j} \hbar}  \sum_{\alpha \beta } \frac{n_{\uu \alpha} - n_{\dd \beta}
     }{\hbar \omega + \epsilon_\alpha - \epsilon_\beta \pm  \iu \delta}
     W_{i \du}^{\alpha \beta} W_{j\ud  }^{\beta \alpha} \, .
\end{align}
This self-energy has equilibrium contributions as well as non-equilibrium contributions, however, the non-equilibrium contributions scale as $\mu_\uu/ \epsilon_F$ and $\mu_\dd / \epsilon_F$ and are disregarded in the following. 
The equilibrium part of \eq \eqref{eq:SigmaRomega} becomes particularly transparent when expanding to first order in the frequency $\omega$:
\begin{align}
    \Sigma_{\ud ij}^{R/A}(\omega)  &\approx \Sigma_{\ud ij}^{R/A}(\omega = 0)
     \pm \iu \omega \alpha_{ij}\, ,
\end{align}
where we introduced the frequency-independent matrix element 
\begin{equation}
\label{eq:beta}
    \alpha_{ij} =  \frac{2 \pi}{\sqrt{s_i s_j}} \sum_{\alpha \beta}  \left[ - f'(\epsilon_\alpha) \right]
    \delta(\epsilon_\alpha - \epsilon_\beta) 
    W_{i \du}^{\alpha \beta} W_{j\ud  }^{\beta \alpha} \, ,
\end{equation}
which can be approximated to 
\begin{equation}
\label{eq:beta2}
    \alpha_{ij} =  \frac{2 \pi}{\sqrt{s_i s_j}} \sum_{\kappa \lambda n m}  
    W_{i \du}^{\kappa n \lambda m} W_{j \ud  }^{\lambda m \kappa n} \, ,
\end{equation}
where the scattering states matrix elements are evaluated at the Fermi surface.
We note from the identity $[W_{i \ud}^{\alpha \beta}]^* = W_{i \du}^{\beta \alpha}$ that $\alpha$ is a Hermitian matrix in the space of lattice sites, i.e. $[\alpha_{ij}]^* = \alpha_{ji}$.
The zeroth order term in frequency is
\begin{align}
    \nonumber
    [S_{22}^{\bar q cl} + S_{22}^{\bar{cl} q}]_0 =& \hbar \sum_{ij}  \int d\omega  \bar d_i^q(\omega)  \Sigma_{\ud ij}^{R}(0) d_{j}^{cl}(\omega) \\
    &+ \hbar \sum_{ij}  \int d\omega \bar d_i^{cl}(\omega)  \Sigma_{\ud ij}^{A}( 0) d_{j}^{q}(\omega) \, ,
\end{align}
which is a constant longitudinal field that plays no role in the instantaneous reference frame, as discussed above.

The first-order term in frequency is finite even in equilibrium,
\begin{align}
    [S_{22}^{\bar q cl} + S_{22}^{\bar{cl} q}]_1 = \hbar \sum_{ij} \alpha_{ij} \int dt \bar D_i \gamma^q \partial_t D_{j}\, ,
\end{align}
and takes the form of a Gilbert damping term, including both inter-lattice and intra-lattice contributions.
The spin transfer torque coefficient $\alpha$ and the spin pumping coefficient $\beta$ are related to each other as a consequence of the Onsager reciprocal relations \cite{brataasSpinPumpingSpin2012a}. In Appendix \ref{app:onsager} we derive this relation, which is given in \eq \eqref{eq:app_alpha=beta}, and derive an optical theorem relating the scattering matrices, given in \eq \eqref{eq:app_optical}.

Summarizing this section, we have found that the corrections to the magnon action $S_m$ in the presence of spin and charge accumulations in surrounding normal metals is $S_1 + S_{21} + S_{22}^{\bar q cl}+ S_{22}^{\bar{cl} q}+ \tilde S_{22}^{qq} + S_{22}^{\bar qq} $, and found that the first three of these contributions appear like magnetic fields and (in the low-frequency limit) like Gilbert-like damping terms in the effective magnon action. 
Importantly, we find both longitudinal and transverse fields in the general case.
The last two contributions to the action consist of coupled quantum fields and are the result of purely quantum effects.
These terms are the subject of the next section.

\subsection{Fluctuating fields}
\label{section:fluctuating}
From the effective action in the last section, we were able to associate the $(q,cl)$ and $(cl, q)$ terms with longitudinal fields by comparing them with the magnon action in \eq \eqref{eq:Smrot}. 
Now, we must address the issue of how to interpret the $(q, q)$ terms, which lack an analog in the action described in \eq \eqref{eq:Smrot}. In this section, we derive fluctuating forces from these terms by employing a Hubbard-Stratonovich (HS) transformation on the quadratic fields in the effective action, introducing auxiliary fields in the process.
Commencing with the contribution from the term $S_{22}^{ \bar q  q}$, we introduce the complex auxiliary field $h^{\bar q q}_i$  (in units of inverse second) via a conventional Hubbard–Stratonovich transformation:
\begin{widetext}
\begin{align}
    \nonumber
    \e^{ \iu  S_{22}^{\bar q q}/\hbar} &= 
    \exp \biggl[\int dt dt' \sum_{ ij }  \bar d^q_i(t) \iu  \Sigma_{ij}^K(t- t') d^{q}_{j}(t') \biggr]
    \\
    &= \frac{1}{ \det{[  - \iu \Sigma^K ]}}
     \int \prod_i  D [ h^{\bar qq}_i]
    \exp \biggl[  \iu \int dt  \sum_i  h^{\bar qq}_i(t) \bar d_i^q(t) + h.c.  -  \int dt dt' \sum_{ij}  \bar h^{\bar qq}_i(t) [- \iu  \Sigma^K_{ ij}(t-t')]^{-1}  h^{qq}_{j}(t') \biggr] \, ,
    \label{eq:HS2}
\end{align}
\end{widetext}
where a shorthand notation for the measure was introduced as $D[h^{\bar qq}_i] = \Pi_k \left\{ d[ \Im h^{\bar qq}_i(t_k)] d [\Re h^{\bar qq}_i(t_k)] / \pi \right\}$, where $k$ is the index used to order the discretization of the time coordinate.
From the Gaussian form of \eq \eqref{eq:HS2}, the correlators of the auxiliary field can be identified as
\begin{subequations}
\begin{align}
    \langle h_i^{ \bar q  q}(t) \rangle &= 0 \\
    \langle h_i^{\bar  q  q}(t)  h_{j}^{ \bar q  q}(t') \rangle &= 0\\
    \langle \bar h_i^{ \bar q  q}(t)  h_{j}^{ \bar q  q}(t') \rangle &= - \iu  \Sigma^K_{ij}(t - t') \, .
\end{align}
\end{subequations}
The second term in the exponent is quadratic in the new fields, and gives no contribution to the magnon action, while the first term is linear in the magnon field $d_i$ and is interpreted as an effective transverse field in the magnon action.

The contribution from the terms $\tilde S_{22}^{qq} + \tilde S_{22}^{\bar q \bar q}$ is HS transformed by performing an unconventional transformation in the two complex fields $\tilde h_i^{qq}$ and $\overline{\tilde h_i^{qq}}$ separately:
\begin{widetext}
\begin{align}
    \nonumber
    \e^{ \iu  \tilde S_{22} / \hbar} &= 
    \frac{1}{\sqrt{\det{[- 2 \iu  \tilde \Sigma_{\du}^K  ]}} \sqrt{\det{[- 2 \iu \tilde \Sigma_\ud^K]}}}
     \int \prod_i  D [\tilde h^{qq}_i]
    \exp \Biggl[  \iu \int dt  \sum_i 
    \overline{
    \begin{pmatrix}
        \tilde h^{qq}(t)  &  \overline{\tilde h^{q q}}(t)
    \end{pmatrix}}_i
    \begin{pmatrix}
        d(t)
        \\
        \bar d(t)
    \end{pmatrix}_i 
    + h.c.
    \\
    & \hspace{9.5em} -  \int dt dt' \sum_{ij} \overline{
    \begin{pmatrix}
        \tilde h^{qq}(t)  &  \overline{\tilde h^{q q}}(t)
    \end{pmatrix}}_i
    \begin{pmatrix}
       0 & - \iu \tilde \Sigma_{\du }^K(t- t') \\
       - \iu \tilde \Sigma_{\ud }^K(t- t')  & 0
   \end{pmatrix}_{ij}^{-1}
    \begin{pmatrix}
        \tilde h^{qq}(t')  \\ \overline{\tilde h^{ q  q}}(t')
    \end{pmatrix}_{j}
    \Biggr]\, ,
    \label{eq:HS}
\end{align}
\end{widetext}
again interpreting the exponent as an effective action including the field $\tilde h_{i}^{qq}$, which has the correlators
\begin{subequations}
    \label{eq:fluc_hqq}
\begin{align}
    \langle \tilde h_i^{ q  q}(t) \rangle &= 0 \\
    \langle \tilde h_i^{ q  q}(t)  \tilde h_{j}^{ q  q}(t') \rangle &= - \iu \tilde \Sigma_{\ud ij}^K(t- t')\\
     \langle \overline{\tilde h_i^{ q  q}}(t)  \overline{\tilde h_{j}^{ q  q}}(t') \rangle &= - \iu \tilde \Sigma_{\du ij}^K(t- t')\\
    \langle \overline{\tilde h_i^{ q  q}}(t)  \tilde h_{j}^{ q  q}(t') \rangle &= 0\, .
\end{align}
\end{subequations}
We remark that the unconventional form of the Hubbard-Stratonovich decoupling leads to non-zero correlators for equal fields, as opposed to the conventional approach where the non-zero correlators involve one field being the complex conjugate of the other.
The fields $h^{\bar qq}$ and $\tilde h^{qq}$ are interpreted as fluctuating transverse fields with, in general, different amplitudes depending on the lattice site, but with correlators between lattice sites. 
Comparing the effective action in \eq \eqref{eq:HS2} and \eq \eqref{eq:HS} with the magnon action in \eq \eqref{eq:Smrot}, the components of the total fluctuating field $\vec H^\text{f}$ can be identified as
\begin{subequations}
\label{eq:fluc:H+}
\begin{align}
    \gamma \mu_0 H^\text{f}_{+, i} &=- \frac{1}{\sqrt {s_i} } \left[ 2 \tilde h^{q q}_i + h^{\bar qq}_i \right]
    \\
     \gamma \mu_0 H^\text{f}_{-, i} &= - \frac{1}{\sqrt {s_i} } \left[2 \overline{\tilde h}^{qq}_i + \bar h^{\bar qq}_i \right] \, .
\end{align}
\end{subequations}
In this expression, the factor of $2$ arises from the unconventional nature of the Hubbard-Stratonovich transformation in \eq \eqref{eq:HS}.
The correlators between the Cartesian components of the fluctuating field can be calculated using \eq \eqref{eq:fluc_hqq}  and \eq \eqref{eq:fluc:H+},
\begin{subequations}
\begin{align}
     2 \sqrt{s_i s_j} \gamma^2 \mu_0^2 \langle H^\text{f}_{x i} H^\text{f}_{x j} \rangle &=   
    \Im \Sigma_{ij}^K
    + 4 \Im \tilde \Sigma_{\ud ij} 
    \label{eq:corrxx}
    \\
    2 \sqrt{s_i s_j} \gamma^2 \mu_0^2 \langle H^\text{f}_{xi}  H^\text{f}_{y j}  \rangle &= - \Re \Sigma_{ij}^K  
    - 4 \Re \tilde \Sigma_{\ud ij}
    \\
    \label{eq:corryy}
    2 \sqrt{s_i s_j} \gamma^2  \mu_0^2 \langle H^\text{f}_{y i} H^\text{f}_{y j} \rangle &= \Im \Sigma_{ij }^K -
     4 \Im \tilde \Sigma_{\ud ij} \, ,
\end{align}
\label{eq:cors}
\end{subequations}
from which we conclude that the correlators in the fluctuating field $\vec H^\text{f}$ are real numbers. In \eq \eqref{eq:cors}, we omitted the time arguments for notational simplicity.
Furthermore, it is evident that for $i=j$ and $t = t'$, the correlators in \eq \eqref{eq:corrxx} and \eqref{eq:corryy} are positive, aligning with the conditions expected for representing the variance of a real field. 

\begin{widetext}
\subsection{Equations of motion}
\label{section:eom}
After HS decoupling the $qq$ components, the effective action reads
\begin{align}
 \nonumber
     S_\text{eff} =&   -  \gamma \hbar \mu_0 \int dt 
     \left[  
     \sum_{a  } 
     \sqrt{s_a}
    \left( H_{a+}^\text{stt} + H^\text{f}_{a+}
    \right) \bar a^{q}_a(t) 
    +
    \sum_{b  } 
    \sqrt{s_b}
    \left( H_{b-}^\text{stt} + H^\text{f}_{b-}
    \right) \bar b^{q}_b(t)
    + h.c.
    \right]
    \\
    &
    + \hbar \int dt  \left[ 
    \sum_{a a'}  \beta_{aa'}\bar a_a^q  \partial_t a_a^{cl}
    +\sum_{a b}  \beta_{ab}   a_a^q  \partial_t  b_b^{cl}
    +\sum_{b a}  \beta_{ba}  \bar b_b^q  \partial_t  \bar a_a^{cl} 
    +\sum_{b b'}  \beta_{bb'}\bar b_b^q    \partial_t b_{b'}^{cl} 
    + h.c. 
    \right] \, .
\end{align}
Having cast the total action $S_m + S_\text{eff}$ in a form that is linear in the quantum fields $\bar a^q$ and $\bar b^q$ and their complex conjugates, we can integrate over these fields in the partition function, producing the functional delta function imposing the semiclassical equations of motion for the fields $a^{cl}$ and $b^{cl}$~\cite{kamenev_field_2011}. Using $a^{cl}_a = S_{+a}/ (\hbar \sqrt{s_a})$ and $\bar b^{cl}_b = S_{+b}/(\hbar \sqrt{s_b})$ in the semiclassical limit, we find the coupled equations of motion:
\begin{align}
\label{eq:eom_S}
    \iu  \partial_t S_{i+} = \hbar^{-1} E_i S_{i+} 
    +  s_i \hbar \mu_0 \gamma \left(  H_{i+} +   H^\text{f}_{i+} + H^\text{stt}_{i +}
 \right)
     - \sum_{j} \beta_{ij} \partial_t S_{j+} \, ,
\end{align}
as well as its complex conjugated counterpart. 
Both in the definition of this field and in \eq \eqref{eq:eom_S}, the upper sign holds for sublattice $\mathcal A$, and the lower sign holds for sublattice $\mathcal B$.
We find the Cartesian components by taking the real and imaginary parts and divide with $\hbar s_i$ to find an equation for the vector $\vec m_i = \vec S_i / (\hbar s_i) $,
\begin{align}
    \partial_t \vec m_{i}  &=  \vec  \tau_{i}^\text{b} +  \vec \tau^\text{f}_{i} + \vec  \tau_{i}^\text{sp} + \vec  \tau_{i}^\text{stt} \, ,
\end{align}
where
\begin{subequations}
\begin{align}
    \vec \tau^\text{b}_i &= - \vec z \times \left(\hbar^{-1} E_i \vec m_{i} +   \gamma \mu_0 \vec H_{i}  \right)
    \\
    \vec \tau_i^\text{f} &= -   \gamma \mu_0 \vec z  \times 
    \vec H^\text{f}_{i}
    \\
    \vec \tau_i^\text{stt} &= -   \gamma \mu_0 \vec z \times \vec H_i^\text{stt}
    \\
    \vec \tau_i^\text{sp} &= \sum_j \Re \beta_{ij} \vec z \times \partial_t \vec m_j + \sum_j \Im \beta_{ij} \vec z \times ( \vec z \times \partial_t \vec  m_j)
\end{align}
\end{subequations}
is microscopic expressions for the bulk torque $\tau^\text{b}$, the fluctuating torque $\vec \tau^\text{f}$, the spin pumping torque $\vec \tau^\text{sp}$, and the spin transfer torque $\vec \tau^\text{stt}$.
\end{widetext}

\section{Conclusion}
\label{section:conclusion}
In this paper, we have presented a general quantum theory of spin dynamics in magnet-normal metal systems, generalizing earlier results to a general antiferromagnetic or ferrimagnetic bipartite lattice. Spin and charge accumulations in the normal metals influence the magnetization dynamics in the magnet through spin transfer torque, and the damping is enhanced due to spin pumping, including both inter- and intra-lattice contributions. 
We derived expressions for transverse fluctuating fields arising due to the electron magnon interactions. These fields have contributions from equilibrium terms as well as charge and spin accumulation in the normal metals. 
We found site-dependent shot noise contributions that are non-negligible at low temperatures.

\acknowledgments
This work was supported by the Research Council of
Norway through its Centers of Excellence funding scheme,
Project No. 262633, ''QuSpin''.

\appendix

\section{Holstein-Primakoff transformation}
\label{appendix:transformations}
In this Appendix, we discuss the transformations used to diagonalize the magnon Hamiltonian of \eq \eqref{Hm}. To go from the SU(2) spin operators to bosonic annihilation and creation operators, we employ the Holstein-Primakoff transformation \cite{holsteinFieldDependenceIntrinsic1940b, rezendeMultimagnonTheoryAntiferromagnetic1976} at sublattices $\mathcal A$ and $\mathcal B$ and expand to the lowest order in the bosonic operators, assuming the antiferromagnet is close to the Néel state, i.e. that all spins on sublattice $\mathcal A$ ($\mathcal B$) is close to being parallel (antiparallel) to the $z$-direction. At sublattice $\mathcal A$, we expand
\begin{align}
\label{Saplus}
\hat S_{a+} &= \hbar \sqrt{2 s_a} \left(1 - \frac{\hat a^\dag_a \hat a_a}{2 s_a} \right)^{1/2} \hat a_a \approx \hbar \sqrt{2 s_a} \hat a_a 
\\
\label{Saminus}
\hat S_{a-} &= \hbar \sqrt{2 s_a} \hat a^\dag_a \left(1 - \frac{\hat a^\dag_a \hat a_a}{2 s_a}\right)^{1/2}  \approx \hbar \sqrt{2 s_a} \hat a^\dag \\
\label{Sz_a}
\hat S_{az} &= \hbar(s_a - \hat a^\dag_a \hat a_a)   \, , 
\end{align}
where $a_a$ annihilates a localized magnon and $s_a$ is the total spin at lattice site $a$.
In the expansion of the square roots in \eq \eqref{Saplus} and \eq \eqref{Saminus}, we assumed $s_a \gg 1$ and expanded the square root to lowest order in $1/s_A$.
We have employed the standard raising and lowering spin operators, defined as $S_\pm = S_x \pm \iu S_y$.

Similarly, at sublattice $\mathcal B$, we expand
\begin{align}
\hat S_{b+} & =  \hbar \sqrt{2 s_b} \hat b_b^\dag \left(1 - \frac{\hat b^\dag_b \hat b_b}{2 s_b} \right)   \approx  \sqrt{2 s_b} \hat b_b^\dag \, , 
\\
\hat S_{b-} & =   \hbar \sqrt{2 s_b} \left(1 - \frac{\hat b^\dag_b \hat b_b}{2 s_b} \right) \hat b_b \approx  \sqrt{2 s_b} \hat  b_b  \, , 
\\
\label{Sz_b}
\hat S_{bz} & = \hbar \left( -s_b + \hat b^\dag_b \hat  b_b \right)   \, , 
\end{align}
where $\hat b$ annihilates a localized spin-up magnon.

\section{Relating spin transfer torque and spin pumping coefficients}
\label{app:onsager}
We relate the spin transfer pumping coefficients defined in \eq \eqref{eq:beta} to the spin transfer coefficients found in \eq \eqref{eq:alpha} in the case of one normal metal reservoir using the Onsager reciprocal relations \cite{brataasSpinPumpingSpin2012a}.
We start by defining the pumped spin current (in units of electrical current, i.e. Ampere) into normal metal as the change in total spin inside the antiferromagnetic due to spin pumping, i.e. 
\begin{align}
    \vec I^S = - \frac{e}{\hbar} \sum_j S_j \vec \tau^\text{sp}_j \, .
\end{align}
The appearance of $S_j= \hbar \sqrt{s_j(s_j+1)}$ is due to the way we have defined the torques in the main text, causing them to have the dimension of inverse time.
The dynamics of the localized magnetic moment $ \vec \mu_j =  -\gamma S_j \vec m_j$ and the spin current are driven by the external effective field $\vec H_\text{eff}$ and the spin accumulation $\vec \mu^S$, which are the thermodynamic forces in our system. In linear response, we can then write the equations for the spin dynamics and the spin current in matrix form:
\begin{align}
\label{eq:linear_response}
    \begin{pmatrix}
        - \gamma S_i \partial_t \vec m_i
        \\
        \vec I^S
    \end{pmatrix}
    =
    \begin{pmatrix}
        L^{mm}_{ij} & L^{ms}_{i}\\
        L^{sm}_{j} & L^{ss}
    \end{pmatrix}
    \begin{pmatrix}
        \mu_0 \vec H^\mathrm{eff}_{j} \\ \vec \mu^S / e
    \end{pmatrix} \, ,
\end{align}
where the matrix elements $3 \times 3$ tensors that effectively apply the relevant cross products to make \eq \eqref{eq:linear_response} consistent with the Landau-Lifshitz equation, and where we use the Einstein summation convention for repeated Latin indices.

\subsection{Identifying \texorpdfstring{$L^{sm}$}{TEXT}}
\label{Lsm}

Inserting the spin pumping torque from \eq \eqref{eq:tau_sp_main}, the spin current becomes
\begin{align}
    \vec I^S = - X_j \partial_t \vec m_j \, ,
\end{align}
where we defined the $3\times 3$ matrix $X_j$ as
\begin{align}
    X_{j} = \frac{e}{\hbar} S_j \sum_i \left[ \alpha_{ij}^R \tilde O
    + \alpha_{ij}^I \tilde O^2 
    \right] \,,
\end{align}
and the $3 \times 3$ matrix $\tilde O$ implements the cross product $\vec z \times \vec v = \tilde O \vec v$ and can be defined in terms of the Levi-Civita tensor.
The LLG equation in the absence of spin accumulation (causing the spin transfer torque to vanish) reads 
\begin{align}
    (1 - \alpha_b \tilde O) \partial_t \vec m_i = \tilde O (- \gamma \mu_0 \vec  H^{\text{eff}}_i),
\end{align}
where $\alpha_b$ is the (bulk) Gilbert damping constant.
Hence, we identify
\begin{align}
\label{eq:Lsm}
    L^{sm}_{j} =  \gamma X_j  \tilde O (1 - \alpha_b \tilde O)^{-1}\, .
\end{align}

\subsection{Identifying \texorpdfstring{$L^{ms}$}{TEXT}}
\label{sect:Lms}
Inserting the spin transfer torque from \eq \eqref{eq:stt_main} into the LLGS equation in the absence of an effective field, we find
\begin{align}
    \partial_t \vec m_i 
    = 
    \hbar^{-1}
    ( 1 - \alpha_b \tilde O)^{-1}  \left[\beta_{i }^I \tilde O  
    - \beta_{i }^R \tilde O^2  \right] \vec \mu^S \, ,
\end{align}
meaning that we can identify the linear response coefficient $L^{ms}$ as (no Einstein summation)
\begin{align}
\label{eq:Lms}
    L^{ms}_{i}= - \frac{S_i \gamma e}{\hbar} ( 1 - \alpha_b \tilde O)^{-1} \left[\beta_{i }^I \tilde O  
    - \beta_{i }^R \tilde O^2  \right] \, .
\end{align}

\subsection{Deriving relations from the Onsager reciprocal relations}
We are now looking to employ Onsager's reciprocal relation:
\begin{align}
\label{eq:ons1}
    [L^{sm}_{i}(\{-\vec m_j\})]^T =  L^{ms}_{i}(\{\vec m_j\}) \, ,
\end{align}
where the superscript $T$ indicates a matrix transpose in the $3\times3$ Cartesian space.
Using the matrix identity $\tilde O^3 = -\tilde O$, we find that \eq \eqref{eq:ons1} implies that
\begin{align}
 \beta_{j }^I \tilde O  
                -\beta_{j }^R \tilde O^2 
    =  \sum_i \left[\alpha_{ij}^I \tilde O  - \alpha_{ij}^R \tilde O^2 
    \right] 
\end{align}
This equality is satisfied if
\begin{align}
\label{eq:app_alpha=beta}
     \beta_{j } =  \sum_i \alpha_{ij} \, ,
\end{align}
which generalizes the result from Ref.\ \onlinecite{Brataas:PRB2022}.
Inserting the definitions of these coefficients in the low-temperature limit, we find that
\begin{align}
\label{eq:app_optical}
    \sum_n    W_{j \ud}^{ n  n}  =  \iu \pi \sum_{ i n m}
    W_{i \du}^{n  m} W_{j \ud  }^{ m  n}
    \, ,
\end{align}
which we classify as a generalized optical theorem, since in the diagonal case $i = j$, we can rewrite the imaginary part of this to 
\begin{align}
\label{eq:app_optical2}
    \Im \left[ \sum_n    W_{i \ud}^{ n  n} \right] =   \pi \sum_{ i n m}
    |W_{i \du}^{n  m}|^2
    \, ,
\end{align}
which is reminiscent of the optical theorem in wave scattering theory.

\section{Contour fields and Keldysh rotations}
\label{appendix:pm}
In this Appendix, we show how the action can be written in the $\pm$ basis, and introduce the Keldysh rotated fields, which differ in the case of fermionic and bosonic fields.
In the $\pm$ field basis, the action of the scattering (electron) states, corresponding to the Hamiltonian in \eq \eqref{H_escat}, reads 
\begin{align}
    \nonumber
   S_e + S_0 =&
    \sum_{s}\int_{-\infty} ^\infty dt
     \bar c^+_{s} ( \iu \hbar  \partial_t - \epsilon) c_{s}^+
     \\
     &
     -
     \sum_{s}\int_{-\infty} ^\infty dt
     \bar c^-_{s} ( \iu \hbar  \partial_t - \epsilon) c_{s}-
   \nonumber
   \\
   =& 
   \sum_{s \xi t}
     \bar c^\xi_{s} ( \iu \hbar  \partial_t - \epsilon) c_{s}^\xi \, ,
    \label{Se}
\end{align}
where now $c_s$ is a vector containing the scattering fields, $\bar c_s$ denotes its complex conjugate, and $\epsilon$ is a diagonal matrix containing all energy eigenvalues of the scattering states.
In the final line, we have written the time integration as a sum for concise notation.
Additionally, we introduced the sum over ''$\pm$'' fields as a sum  over $\xi = \{+, -\}$, with an implicit negative sign before the ''-'' field, i.e. $\sum_\xi \ldots^\xi = \ldots^+ - \ldots^-$.
A similar notation will also be used for the magnon fields below.
The negative sign ($\xi = -$) in the integral in \eq \eqref{Se} and in the other actions below originates from reversing the integration limits on the backward contour.
The magnon action is
\begin{align}
\label{Sm}
\nonumber
    S_m    =& 
     \sum_{\xi ab t} [\bar  a_a^\xi( \iu  \hbar \partial_t - E^A_{ab})  a_a^\xi  
     + \bar b_b^\xi ( \iu \hbar \partial_t - E^B_{ab})    b_b^\xi]
    \\
    \nonumber
    &
    -
    2 \sum_{aa'}J_{aa'} \sqrt{s_a s_{a'}} \bar a^\xi_a  a_{a'}^\xi 
    \\ \nonumber
    &
    -
    2 \sum_{bb'} J_{bb'} \sqrt{s_b s_{b'}} \bar b^\xi_b  b^\xi_{b'}
    \nonumber
    \\ \nonumber
    &
    -
    2 \sum_{\xi ab t} J_{ab}   \sqrt{s_a s_b}
    [a_a^\xi  b_b^\xi +\bar a_a^\xi  \bar b_b^\xi]
    \\
    \nonumber
    &
    -  \gamma \mu_0 \hbar \sum_{\xi a t} \sqrt{\frac{s_a}{2}} [ H^A_{a-}  a_a^\xi + H^A_{a+} \bar a_a^\xi]
    \\    &
    -  \gamma \mu_0 \hbar \sum_{\xi b t} \sqrt{\frac{s_b}{2}}   [ H^B_{b-} \bar b_b^\xi + H^B_{b+} b_b^\xi ]\, .
\end{align}
The first-order electron-magnon interaction is
\begin{align}
        S_1 =& -  \sum_{\substack{\xi a t \\ \alpha \beta} }  
        \sqrt{\frac{2}{s_a}}
        \left[  {a}_a^\xi  \bar{c}_{\downarrow \alpha}^\xi  W_{a \downarrow \uparrow}^{\alpha \beta}{c}_{\uparrow \beta}^\xi + \bar{a}_a^\xi  \bar{c}_{\uparrow \alpha}^\xi  W_{a\uparrow \downarrow}^{\alpha \beta} c_{\downarrow \beta}^\xi \right]
        \nonumber
        \\
        &  
        -
         \sum_{\substack{\xi b t \\ \alpha \beta}} 
         \sqrt{\frac{2}{s_b}} 
        \left[  \bar{b}_b^\xi  \bar{c}_{\downarrow \alpha}^\xi  W_{b \downarrow \uparrow}^{\alpha \beta} c_{\uparrow \beta}^\xi + b_b^\xi  \bar{c}_{\uparrow \alpha}^\xi  W_{b\uparrow \downarrow}^{\alpha \beta} c_{\downarrow \beta}^\xi \right] \, ,
        \label{S1}
\end{align}
and the second-order term is
\begin{align}
    \nonumber
    S_2 =& 
     \sum_{\substack{\xi a t \\ \alpha \beta}}
     \frac{1}{s_a}
     \bar{a}_a^\xi {a}_a^\xi  
     \left[ \bar{c}_{\uparrow \alpha}^\xi  W_{a\uparrow \uparrow}^{\alpha \beta} {c}_{\uparrow \beta}^\xi - \bar{c}_{\downarrow \alpha}^\xi W^{ \alpha \beta}_{a \downarrow \downarrow} {c}_{\downarrow \beta}^\xi  
    \right] 
    \\ 
    & - \sum_{\substack{\xi b t \\ \alpha \beta}} 
    \frac{1}{s_b} 
    \bar{b}_b^\xi {b}_b^\xi  \left[ \bar{c}_{\uparrow \alpha}^\xi W_{b \uparrow \uparrow}^{\alpha \beta} {c}_{\uparrow \beta}^\xi - \bar{c}_{\downarrow \alpha}^\xi W^{ \alpha \beta}_{b \downarrow \downarrow} {c}_{\downarrow \beta}^\xi  \right] \, . 
\end{align}

For a general bosonic field $\phi$, the classical ($cl$) and quantum ($q$) fields are defined as \cite{kamenev_field_2011}:
\begin{align}
\label{eq:clq_ief}
    \phi^{cl/q} = \frac{1}{\sqrt{2}} (\phi^+ \pm \phi^-) && \bar \phi^{cl/q} = \frac{1}{\sqrt{2}} (\bar \phi^+ \pm  \bar\phi^-)\, .
\end{align}
In our case, we have $\phi  = \{a, b\}$.
The upper (lower) sign holds for the classical (quantum) fields. For a fermionic field $c$, the rotated fields are denoted by $1$ and $2$, and defined as
\begin{align}
    c^{1/2} = \frac{1}{\sqrt{2}} (c^+ \pm c^-) && \bar c^{1/2} = \frac{1}{\sqrt{2}} (\bar c^+ \mp \bar c^-) \, .
\end{align}
For fermions, $\bar c$ and $c$ are independent variables, not related by complex conjugation.

\section{Fourier transform}
\label{appendix:fourier}
For a general function of relative time $t-t'$, we define the Fourier transform between the relative time domain and the energy domain as
\begin{align}
    f(\omega) &= \int_{-\infty}^\infty d(t- t') \e^{ \iu \omega (t - t')} f(t- t') \, ,
    \\
    f(t - t') &= \int_{-\infty}^\infty  \frac{d\omega}{2 \pi} \e^{- \iu \omega (t - t')} f(\omega)\, .
\end{align}
The delta function can be represented as
\begin{align}
    \delta(t - t') &= \int_{-\infty}^\infty  \frac{d\omega}{2 \pi} \e^{- \iu \omega (t - t')} \, , 
    \\
    \delta(\omega) &= \frac{1}{2 \pi} \int_{-\infty}^\infty d(t- t') \e^{ \iu \omega (t - t')} \, .
\end{align}
Finally, we note a frequently employed identity,  
\begin{align}
    \label{eq:ap:heaviside}
    -\iu \int_{-\infty}^\infty d(t- t') \e^{ \iu \omega (t - t')}  \theta(t - t') &= (\omega  + \iu \delta)^{-1} \, , 
    \\
    \iu \int_{-\infty}^\infty d(t- t') \e^{ \iu \omega (t - t')} \theta( t'-t ) &= (\omega -   \iu \delta)^{-1} \, ,
\end{align}
where $\delta$ is an infinitesimal positive quantity.

\bibliography{bibliography.bib}

\end{document}